\documentclass[conference]{IEEEtran}
\IEEEoverridecommandlockouts

\usepackage[table]{xcolor}      
\usepackage{booktabs}           
\usepackage{tabularx}           
\usepackage{array}              
\usepackage{colortbl}           
\usepackage{fontawesome5}       
\usepackage{amsmath,amssymb}    
\usepackage{mathrsfs}           
\usepackage{microtype}          
\usepackage{float}

\definecolor{quantumBlue}{RGB}{25,118,210}    
\definecolor{cryptoGreen}{RGB}{56,142,60}     
\definecolor{mathAmber}{RGB}{230,126,34}      
\definecolor{systemPurple}{RGB}{123,31,162}   
\definecolor{futureTeal}{RGB}{0,137,123}      
\definecolor{headerGray}{RGB}{69,90,100}      
\definecolor{quantumBlueLight}{RGB}{232,240,254}
\definecolor{cryptoGreenLight}{RGB}{232,245,233}
\definecolor{mathAmberLight}{RGB}{255,243,224}
\definecolor{systemPurpleLight}{RGB}{243,229,245}
\definecolor{futureTealLight}{RGB}{224,242,241}

\newcolumntype{L}[1]{>{\raggedright\arraybackslash}p{#1}}

\usepackage{cite}
\usepackage{amsmath,amssymb,amsfonts}
\usepackage{algorithmic}
\usepackage{graphicx}
\usepackage{textcomp}

\usepackage[table,dvipsnames]{xcolor}
\usepackage{booktabs}
\usepackage{multirow}
\usepackage{tabularx}
\usepackage{colortbl}
\usepackage{array}

\usepackage{tikz}
\usetikzlibrary{arrows.meta,positioning,fit,backgrounds,
                shapes.geometric,shadows,calc}
\usepackage{wasysym}   
\usepackage{url}
\usepackage[hidelinks]{hyperref}
\usepackage{microtype}
\usepackage{mathrsfs}   
\usepackage{bm}         
\usepackage{nicefrac}   

\newcommand{\api}[1]{\ensuremath{\mathsf{#1}}}          
\newcommand{\val}[1]{\ensuremath{\mathit{#1}}}           
\newcommand{\ep}[1]{\textsl{\textsf{#1}}}               
\newcommand{\badge}[2]{\ensuremath{\mathcal{Q}_{#1}\,|\,\mathtt{#2}}}  
\newcommand{\qnum}{\ensuremath{q_{\mathrm{num}}}}        
\newcommand{\bellvec}{\ensuremath{\boldsymbol{\beta}}}   
\newcommand{\pkh}{\ensuremath{\mathcal{H}_{\mathrm{pk}}}} 

\definecolor{qblue}{RGB}{30,100,200}
\definecolor{qpurple}{RGB}{120,40,180}
\definecolor{qgreen}{RGB}{20,140,80}
\definecolor{qorange}{RGB}{210,100,20}
\definecolor{qgray}{RGB}{90,90,100}
\definecolor{qred}{RGB}{180,30,30}
\definecolor{rowA}{RGB}{240,245,255}   
\definecolor{rowB}{RGB}{245,240,255}   
\definecolor{rowC}{RGB}{240,250,243}   
\definecolor{hdrblue}{RGB}{30,100,200} 
\definecolor{hdrpurple}{RGB}{120,40,180}
\definecolor{hdrgreen}{RGB}{20,140,80}
\colorlet{qscicomm}{black!55}   

\begin{document}

\title{QSignAI: Quantum-Randomness-Seeded Identity Signatures at the Intersection of AI for Science and Science for AI}

\author{

  \IEEEauthorblockN{Dongping Liu\textsuperscript{\dag}\textsuperscript{\ddag}}
  \IEEEauthorblockA{\textit{Amazon Web Services}\\
    Hong Kong, China\\}
  \and
  \IEEEauthorblockN{Aoyu Zhang\textsuperscript{\dag}}
  \IEEEauthorblockA{\textit{Amazon Web Services}\\
    Beijing, China\\}
  \and
  \IEEEauthorblockN{Luyao Zhang\textsuperscript{\dag}\textsuperscript{*}}
  \IEEEauthorblockA{\textit{Duke Kunshan University}\\
    Suzhou, China\\}
  \thanks{\textsuperscript{*}Corresponding author: Luyao Zhang (lz183@duke.edu), Digital Innovation Research Center and Social Science Division, Duke Kunshan University. Address: Duke Avenue No.8, Kunshan, Suzhou, Jiangsu, China, 215316. \textsuperscript{\dag} Equal contributions. Authors are listed in alphabetical order by last name and then first name.  \textsuperscript{\ddag} Work done while at Amazon Web Services. Dongping Liu is currently with Tenorshare, Hong Kong, China.}
}

\maketitle

\begin{figure*}[t]
  \centering

\begin{tikzpicture}[
  font=\sffamily,
  >=Stealth,
  every node/.style={inner sep=0pt, outer sep=0pt}
]

\node[
  rounded corners=5pt, draw=qblue, fill=qblue!10,
  line width=1.4pt,
  minimum width=14.8cm, minimum height=1.1cm,
  align=center, text width=14.4cm, inner sep=6pt
] (part) at (7.5, 5.4) {
  {\normalsize\bfseries\color{qblue}%
   $\bigstar$\enspace Participation Layer}%
  \hfill
  {\small\color{qblue!75!black}%
   $\rightarrow$~Bot-enabled messaging platform
   (Telegram, 1B+ MAU)
   $\cdot$ \textit{\textsf{@mention}} $\to$ Webhook $\to$ \textsc{QSignAI} API}
};

\node[
  rounded corners=5pt, draw=qgray, fill=qgray!8,
  line width=1.4pt,
  minimum width=14.8cm, minimum height=1.1cm,
  align=center, text width=14.4cm, inner sep=6pt
] (bot) at (7.5, 3.5) {
  {\normalsize\bfseries\color{qgray}%
   $\star$\enspace Bot / API Layer}%
  \hfill
  {\small\color{qgray!80!black}%
   $\square$~ECS Fargate
   $\cdot$ $\checkmark$~Webhook validation
   $\cdot$ $\rightarrow$~Mention detection
   $\cdot$ $\blacksquare$~Photo pipeline
   $\cdot$ \textit{\textsf{signatureStatus\,=\,generating}}}
};


\node[
  rounded corners=5pt, draw=qpurple, fill=qpurple!10,
  line width=1.4pt,
  minimum width=4.7cm, minimum height=2.5cm,
  align=center, text width=4.4cm, inner sep=6pt
] (quantum) at (2.5, 0.3) {
  {\small\bfseries\color{qpurple}%
   $\bigstar$\enspace Quantum Layer}\\[4pt]
  {\footnotesize\color{qpurple!80!black}%
   \textbf{AWS Braket SV1 + DM1}\\[2pt]
   $\triangleright$~Ckt A: Two-source QRNG\\[1pt]
   $\triangleright$~Toeplitz extractor\\[1pt]
   $\triangleright$~Ckt B: Bell state $|\Phi^+\rangle$\\[1pt]
   $\triangleright$~ToyLWE signature}
};

\node[
  rounded corners=5pt, draw=qgreen, fill=qgreen!10,
  line width=1.4pt,
  minimum width=4.4cm, minimum height=2.4cm,
  align=center, text width=4.1cm, inner sep=6pt
] (db) at (7.5, 0.3) {
  {\small\bfseries\color{qgreen}%
   $\blacklozenge$\enspace Data Layer}\\[4pt]
  {\footnotesize\color{qgreen!80!black}%
   \textbf{DynamoDB} \& \textbf{S3}\\[2pt]
   $\triangleright$~Messages, positions\\[1pt]
   $\triangleright$~Quantum signatures\\[1pt]
   $\triangleright$~Photos, Braket results}
};

\node[
  rounded corners=5pt, draw=qorange, fill=qorange!10,
  line width=1.4pt,
  minimum width=4.4cm, minimum height=2.4cm,
  align=center, text width=4.1cm, inner sep=6pt
] (ui) at (12.5, 0.3) {
  {\small\bfseries\color{qorange}%
   $\circ$\enspace Presentation Layer}\\[4pt]
  {\footnotesize\color{qorange!80!black}%
   \textbf{CloudFront} CDN\\[2pt]
   $\triangleright$~Next.js wall\\[1pt]
   $\triangleright$~5-second polling\\[1pt]
   $\triangleright$~Drag-and-drop}
};

\draw[->, line width=1.3pt, color=qblue]
  (part.south) -- (bot.north)
  node[midway, right=5pt, font=\footnotesize\sffamily,
       color=qblue, fill=white, inner sep=2pt,
       rounded corners=2pt, draw=qblue!30]
  {\lightning~HTTPS webhook};

\coordinate (busL) at (2.5,  2.2);
\coordinate (busC) at (7.5,  2.2);
\coordinate (busR) at (12.5, 2.2);

\draw[line width=1.0pt, color=qgray] (bot.south) -- (busC);
\draw[line width=1.0pt, color=qgray] (busL) -- (busR);
\draw[->, line width=1.0pt, color=qgray] (busL) -- (quantum.north);
\draw[->, line width=1.0pt, color=qgray] (busC) -- (db.north);
\draw[->, line width=1.0pt, color=qgray] (busR) -- (ui.north);

\node[font=\footnotesize\sffamily, color=qgray,
      fill=white, inner sep=2pt, draw=qgray!35,
      rounded corners=2pt]
  at (7.5, 2.2)
  {\textit{async dispatch} $\cdot$ \textit{read-write}};

\draw[->, line width=1.0pt, color=qpurple]
  (4.65, 1.48) -- (5.10, 1.48)
  node[midway, above=3pt, font=\footnotesize\sffamily,
       color=qpurple, fill=white, inner sep=2pt,
       rounded corners=2pt, draw=qpurple!30]
  {$\bigstar$~sig. + nonce};

\draw[->, line width=1.0pt, color=qgreen]
  (9.70, 1.48) -- (10.30, 1.48)
  node[midway, above=3pt, font=\footnotesize\sffamily,
       color=qgreen, fill=white, inner sep=2pt,
       rounded corners=2pt, draw=qgreen!30]
  {$\blacklozenge$~signed URLs};

\node[font=\small\bfseries\sffamily, color=qpurple,
      rotate=90, align=center]
  at (-0.6, 3.0) {Science $\to$ AI};

\node[font=\small\bfseries\sffamily, color=qorange,
      rotate=90, align=center]
  at (15.6, 3.0) {AI $\to$ Science};

\end{tikzpicture}
  \caption{System architecture of QSignAI showing the bidirectional
    relationship between AI and quantum science.
    \textbf{Layers:}
    {\color{qblue}$\blacksquare$}~\textbf{\color{qblue}Participation}
    (bot-enabled messaging, Telegram 1B+~MAU);
    {\color{qgray}$\blacksquare$}~\textbf{\color{qgray}Bot/API}
    (ECS Fargate, webhook, photo pipeline);
    {\color{qpurple}$\blacksquare$}~\textbf{\color{qpurple}Quantum}
    (SV1 + DM1 $\to$ Toeplitz extractor $\to$ ToyLWE signatures);
    {\color{qgreen}$\blacksquare$}~\textbf{\color{qgreen}Data}
    (DynamoDB + S3);
    {\color{qorange}$\blacksquare$}~\textbf{\color{qorange}Presentation}
    (CloudFront, Next.js wall).
    \textbf{Directions:}
    {\color{qpurple}Sci.$\to$AI} (quantum randomness strengthens identity);
    {\color{qorange}AI$\to$Sci.} (bot makes quantum legible to audiences).}
  \label{fig:architecture}
\end{figure*}
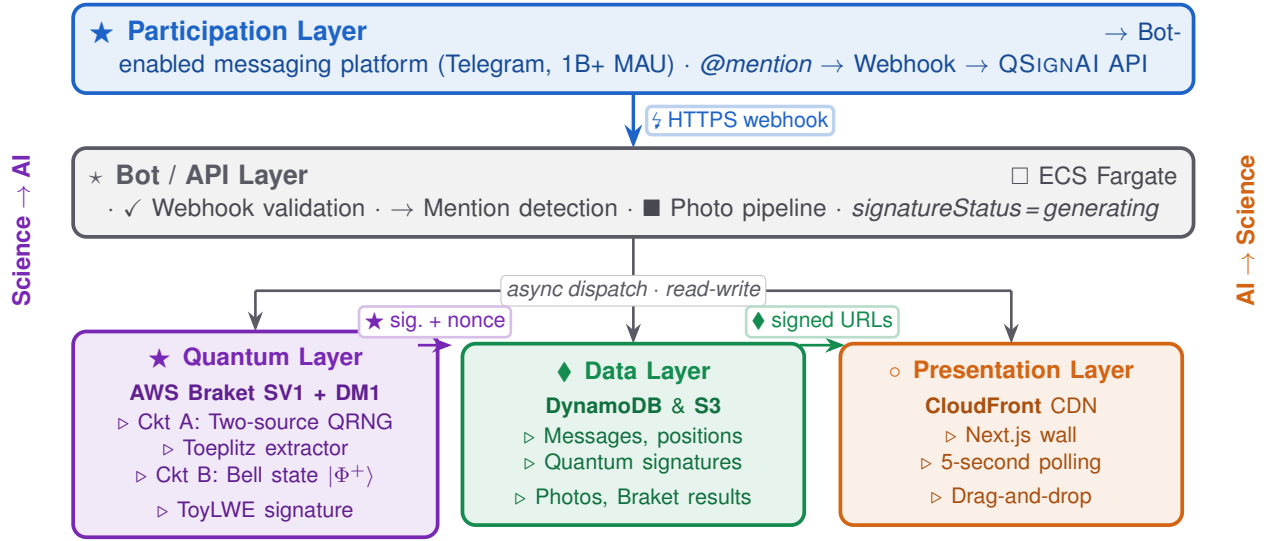
\begin{abstract}
The 2024--2025 Nobel and Turing awards recognised artificial intelligence and quantum science in the same breath --- machine learning as a physical science, artificial intelligence solving 50-year scientific problems, superconducting quantum circuits as the hardware foundation of quantum computing, and quantum information principles as computing's highest achievement. Yet no deployed artificial intelligence system has brought these two streams together for the general public: identity systems still rely on pseudo-random tokens, and quantum circuits remain invisible to the billions of people who use bot-enabled social messaging platforms daily. This paper presents QSignAI, a production-deployed open-source platform demonstrating a bidirectional relationship between artificial intelligence and quantum science in a real-time event participation system. We address three research questions: first, can quantum-randomness generation via independent quantum measurements condensed by a two-source extractor be embedded in an AI-driven social platform with acceptable latency and cost; second, can an AI bot make quantum phenomena perceptually legible to general audiences with no prior technical knowledge; and third, does a system combining both directions work in practice? A conversational AI bot routes each participant's first message through a quantum pipeline comprising a Toeplitz two-source extractor over independent single-qubit Hadamard measurements on SV1 and DM1 simulators, plus a 2-qubit Bell state, producing a unique quantum-randomness-seeded identity signature per participant. The current deployment uses cloud quantum simulators; physical quantum randomness from a quantum processing unit is the near-term extension. The first two research questions are answered through system architecture and qualitative deployment evidence from live events; the third through successful production deployment. Measurable latency benchmarks and controlled user studies are identified as priority future work.
\end{abstract}


\section{Introduction}

In 2024--2025, the Nobel and Turing committees recognised artificial
intelligence~(AI) and quantum science simultaneously. The 2024 Nobel
Prize in Physics went to Hopfield and Hinton~\cite{hopfield1982} for
foundational discoveries enabling machine learning with artificial neural
networks --- the first time the Nobel Committee in Physics recognised AI
research. The 2024 Nobel Prize in Chemistry recognised Hassabis, Jumper,
and Baker~\cite{jumper2021} for AlphaFold2, the canonical demonstration
of AI for Science. The 2025 Nobel Prize in Physics recognised Clarke,
Devoret, and Martinis~\cite{martinis1985} for macroscopic quantum
tunnelling and energy quantisation in electrical circuits --- the direct
physical foundation of today's superconducting quantum computers. The
2025 ACM Turing Award --- computing's highest honour --- recognised
Bennett and Brassard~\cite{bennett1984} for establishing the foundations
of quantum information science, including the principle that quantum
measurement produces randomness no classical adversary can reproduce.

Despite this convergence, three gaps persist. \textit{Gap~1 (Science for
AI):} AI identity systems still use pseudo-random number generators
(PRNGs) --- deterministic algorithms that are theoretically reversible.
Bennett and Brassard's Turing Award-winning insight has not been applied
to AI participation systems. \textit{Gap~2 (AI for Science):} quantum
circuits remain invisible to non-specialists. AlphaFold2 showed AI can
make science accessible at scale~\cite{jumper2021}; can an AI bot do the
same for quantum science? \textit{Gap~3 (AI for Better Life):}
transformation only reaches people when it is accessible. A bot on a
platform billions already use --- zero install, zero technical knowledge
--- is a concrete answer to how quantum science improves everyday life.

This paper addresses three Research Questions~(RQs):
\begin{itemize}
  \item \textbf{RQ1} \textit{(Science for AI):} Can quantum-randomness generation via independent quantum measurement sources condensed by a two-source extractor be embedded in an AI-driven social participation system with acceptable latency, cost, and reliability?
  \item \textbf{RQ2} \textit{(AI for Science):} Can an AI bot make quantum phenomena --- superposition, entanglement, Bell state measurement --- perceptually legible to general audiences with no prior technical knowledge?
  \item \textbf{RQ3} \textit{(Deployment):} Does a system combining both directions work in practice, and what does successful deployment demonstrate about the bidirectional relationship?
\end{itemize}

The contributions of this paper are as follows. We present a complete, production-deployed system demonstrating the bidirectional AI--quantum relationship, including quantum circuit design comprising a two-source random number generator (QRNG) based on independent single-qubit Hadamard measurements on SV1 and DM1 simulators, plus a 2-qubit Bell state on Amazon Web Services~(AWS) Braket SV1 (state-vector simulator), seeding ToyLWE (Toy Learning-With-Errors) identity signatures. We describe the Telegram Bot API integration architecture with multi-group isolation and graceful degradation, and a three-surface User Experience~(UX) making quantum-derived identity legible to general audiences. The open-source implementation is available on GitHub\footnote{https://github.com/QuantBlockchain/QSignAI}.

The present work focuses on production deployment, post-quantum cryptographic identity binding, and two-source extractor theory; the companion paper \emph{Quantum Futures Interactive}~\cite{quantumfutures2026} emphasises educational game design, player-experience evaluation, and museum-scale interactive installations. Both share the AWS Braket and serverless infrastructure stack but have distinct contributions. Complementary work in this broader research program evaluates
visual AI agents for quantum code generation under cost constraints
\cite{liu2026quantumcircuitvisioncostaware} and explores
quantum-computing hardware through interactive cinematic experiences
based on generative world models
\cite{zhang2026quantumcinemainteractivecinematic}.

\section{Background and Related Work}

\subsection{Quantum Randomness}

Classical PRNGs are deterministic: given the same seed, they always
produce the same sequence. A quantum measurement is governed by the Born
rule --- outcomes are determined by probability amplitudes, not hidden
variables, as experimentally confirmed by Aspect et al.~\cite{aspect1982}
closing the Bell inequality loophole established by Bell~\cite{bell1964}.
Cloud-accessible quantum randomness is now available via services such as
AWS Braket. NIST SP~800-90B~\cite{nist90b} defines randomness
requirements for cryptographic applications. The key distinction in this
paper: quantum randomness is the \emph{input} (what the circuits
produce); identity signatures are the \emph{output} (what ToyLWE
derives).

\subsection{Post-Quantum-Inspired Signatures}

The NIST post-quantum cryptography~(PQC) standardisation
process~\cite{nistpqc2022} produced CRYSTALS-Dilithium (FIPS
204~\cite{fips204}), based on the Learning With Errors~(LWE) problem
introduced by Regev~\cite{regev2009}. LWE mixes a secret vector with a
noise vector, producing a distribution computationally indistinguishable
from uniform. ToyLWE (Toy Learning-With-Errors) as used in this work is a demonstrative, pedagogical
instantiation --- structural analogy to LWE, not a production primitive.
The path to production is replacement with CRYSTALS-Dilithium (FIPS~204, ML-DSA)~\cite{fips204} using the same quantum-randomness-seeded entropy pipeline

\subsection{Bot-Enabled Social Messaging Platforms}

Bot-enabled group messaging platforms expose a programmable webhook
application programming interface~(API) enabling AI agents to mediate
between human participants and backend systems. The current deployment
uses Telegram, which surpassed 1~billion monthly active users~(MAU) in
March 2025~\cite{durov2025}. Prior bot applications span education,
healthcare, and civic engagement~\cite{grudin1994}, but none combine
quantum circuits with real-time social display. Existing interactive
event platforms (Slido, Mentimeter) use PRNG tokens with no quantum
provenance.

\subsection{AI for Science}

The AlphaFold2 result~\cite{jumper2021} established the template for AI
for Science: an AI system making a scientific problem accessible at a
scale classical methods cannot match. QSignAI applies the same template
to quantum science communication: an AI bot making quantum phenomena
accessible to a live audience of non-specialists. Quantum machine
learning~\cite{biamonte2017} and Noisy Intermediate-Scale Quantum~(NISQ)
Era computing~\cite{preskill2018} provide the broader context for near-term quantum applications. Recent reviews map how AI supports the quantum-computing stack and
the representation and characterization of quantum systems
\cite{alexeev2025artificial,du2026artificial}; QSignAI complements
this predominantly engineering-oriented literature through
public-facing quantum communication and a QPU-ready
quantum-randomness pipeline for AI-mediated identity.

\section{System Architecture}

The overall architecture is shown in Fig.~\ref{fig:architecture}. The
system comprises three integrated layers.

The \textbf{Participation Layer} is a bot-enabled group messaging
platform. The current deployment uses Telegram's Bot
API\footnote{Telegram Bot API: \url{https://core.telegram.org/bots/api}}
with webhook delivery validated by
\api{X\text{-}Telegram\text{-}Bot\text{-}Api\text{-}Secret\text{-}Token}. Mention detection parses
\api{message.entities} for \api{type{:}\ "mention"} with
offset/length extraction; \api{caption\_entities} handles photo
messages. Photos are downloaded via the platform file API (20~MB limit)
and uploaded to a private, encrypted Amazon Simple Storage Service~(S3) bucket.\footnote{AWS S3: \url{https://docs.aws.amazon.com/s3/}} The
\textbf{Quantum Identity Layer} runs on AWS Braket
SV1\footnote{AWS Braket: \url{https://docs.aws.amazon.com/braket/}} and
is described in Section~\ref{sec:quantum}. The \textbf{Presentation
Layer} is a Next.js\footnote{Next.js: \url{https://nextjs.org/}}
application served via Amazon CloudFront, polling for new messages every
five seconds.

\textit{Infrastructure.} The full stack is defined as
infrastructure-as-code using AWS Cloud Development Kit~(CDK).\footnote{AWS CDK: \url{https://docs.aws.amazon.com/cdk/}} Elastic
Container Service~(ECS) Fargate runs 1--4 container instances,
auto-scaling at 70\% CPU utilisation. Amazon DynamoDB~(DynamoDB) stores messages with
composite key \api{GROUP\#\{groupId\}} /
\api{MSG\#\{timestamp\}\#\{messageId\}}, encrypted at rest with
point-in-time recovery enabled. All credentials are stored in AWS Secrets
Manager; no secrets appear in code or container images. Scaling
parameters are summarised in Table~\ref{tab:fargate}.

\textit{Security.} Multi-layer security includes CloudFront
secret-header validation blocking direct Application Load Balancer~(ALB)
access, Telegram token validation on every webhook request, bearer token
authentication for admin endpoints, HTTP Strict Transport Security~(HSTS)
and Content Security Policy~(CSP) headers, least-privilege Identity and
Access Management~(IAM) roles, and HTML entity encoding for all user
input (4096-character maximum). The system API endpoints are listed in
Table~\ref{tab:api}.

\begin{table}[t]
  \caption{QSignAI REST API Endpoints. Icons denote role:
    $\bigstar$~ingest, $\blacklozenge$~display, $\times$~delete,
    $\circ$~layout, $\bullet$~admin, $\triangleright$~health.}
  \label{tab:api}
  \centering\small
  \setlength{\tabcolsep}{4pt}
  \begin{tabularx}{\columnwidth}{>{\bfseries}lXll}
    \toprule
    \rowcolor{hdrblue}
    \textcolor{white}{Method} &
    \textcolor{white}{Endpoint} &
    \textcolor{white}{Auth} &
    \textcolor{white}{Role} \\
    \midrule
    \rowcolor{qblue!15}
    POST   & \ep{/api/webhook/[gId]}  & Token  & $\bigstar$~Ingest \\
    \rowcolor{qgreen!12}
    GET    & \ep{/api/messages/[gId]} & Public & $\blacklozenge$~Fetch wall \\
    \rowcolor{qred!12}
    DELETE & \ep{/api/messages/[gId]} & Bearer & $\times$~Soft-delete \\
    \rowcolor{qorange!12}
    PATCH  & \ep{/api/messages/[gId]} & Bearer & $\circ$~Position \\
    \rowcolor{qgreen!12}
    GET    & \ep{/api/groups}         & Public & $\star$~List groups \\
    \rowcolor{qblue!15}
    POST   & \ep{/api/admin}          & Passwd & $\bullet$~Admin login \\
    \rowcolor{qgreen!12}
    GET    & \ep{/api/health}         & Public & $\triangleright$~Health \\
    \bottomrule
    \multicolumn{4}{p{\dimexpr\columnwidth-8pt\relax}}{%
      \footnotesize\itshape
      Row color encodes HTTP verb: {\color{qblue}$\blacksquare$}~POST (write),
      {\color{qgreen}$\blacksquare$}~GET (read),
      {\color{qred}$\blacksquare$}~DELETE,
      {\color{qorange}$\blacksquare$}~PATCH (update).
      All via CloudFront TLS\@. \ep{gId}~=~\ep{groupId}.}
  \end{tabularx}
\end{table}

\begin{table}[t]
  \caption{ECS Fargate Auto-Scaling Parameters.
    Value colors encode scale:
    {\color{qgreen}$\blacksquare$}~minimal/stable,
    {\color{qorange}$\blacksquare$}~moderate.}
  \label{tab:fargate}
  \centering\small
  \setlength{\tabcolsep}{4pt}
  \begin{tabularx}{\columnwidth}{lp{1.4cm}X}
    \toprule
    \rowcolor{hdrgreen}
    \textcolor{white}{Parameter} &
    \textcolor{white}{Value} &
    \textcolor{white}{Rationale} \\
    \midrule
    Min instances      & \cellcolor{qgreen!15}1         & Always-on for webhook \\
    Max instances      & \cellcolor{qorange!15}4         & Handle event spikes \\
    CPU target         & \cellcolor{qorange!15}70\%      & Scale before saturation \\
    Scale-out cooldown & \cellcolor{qgreen!15}30~s       & Respond to traffic bursts \\
    Scale-in cooldown  & \cellcolor{qorange!15}60~s      & Avoid flapping \\
    Memory             & \cellcolor{qorange!15}1024~MB   & Next.js + image processing \\
    CPU                & \cellcolor{qgreen!15}0.5~vCPU   & Sufficient for I/O workload \\
    \bottomrule
    \multicolumn{3}{p{\dimexpr\columnwidth-8pt\relax}}{%
      \footnotesize\itshape
      Managed by AWS Application Auto Scaling.
      Min.\ one instance keeps webhook always available.}
  \end{tabularx}
\end{table}

\section{Quantum Randomness Pipeline}
\label{sec:quantum}

\subsection{Why Quantum Randomness}

A PRNG is a deterministic function: given the same seed, it always
produces the same output. An adversary who recovers the seed can reproduce
every identity token the system has ever issued. A quantum measurement has
no such algorithm underneath it. The Born rule states that measurement
outcomes are governed by probability amplitudes --- not hidden variables,
not prior system state. This was experimentally confirmed by Aspect et
al.~\cite{aspect1982}. Quantum-randomness-seeded identity tokens on a physical QPU cannot be predicted or reproduced by any classical adversary, regardless of computational power, because the randomness was not computed --- it was measured~\cite{bennett1984,herrerocollantes2017}. The present deployment uses classical simulators (SV1 and DM1), which are deterministic at the hardware level; the architecture is QPU-ready and the extractor pipeline applies directly to hardware-sourced randomness~\cite{tamura2020}.

\subsection{Circuit A: Two-Source Quantum RNG}

The two quantum circuits are shown in Fig.~\ref{fig:circuits}. Circuit~A
implements a two-source quantum random number generator~(QRNG) on AWS Braket.
Two independent single-qubit Hadamard circuits are executed in parallel:
one on the SV1 state-vector simulator (the \emph{ideal} source) and one on
the DM1 density-matrix simulator (the \emph{noisy} source). Each circuit
applies a Hadamard gate to place the qubit in superposition, then measures
in the computational basis. Repeating this process $n$ times yields two
raw bit strings:
\begin{itemize}
  \item $\mathbf{X} \in \{0,1\}^n$ from SV1, representing the ideal
    quantum randomness, and
  \item $\mathbf{Y} \in \{0,1\}^n$ from DM1, capturing physically noisy
    but independently generated randomness.
\end{itemize}
These raw streams are not individually uniform---SV1 is a deterministic
simulator and a single-source modal outcome is predictable---so they are
condensed through a Toeplitz two-source extractor. Let $T$ be an
$m \times n$ binary Toeplitz matrix whose first row and first column are
derived from $\mathbf{Y}$. The extractor computes
\begin{equation}
  \mathbf{Z} \;=\; \mathrm{Ext}(\mathbf{X}, \mathbf{Y})
       \;=\; T \cdot \mathbf{X} \;\oplus\; \mathbf{g},
  \label{eq:toeplitz}
\end{equation}
where $\mathbf{g} \in \{0,1\}^m$ is a universal hash offset also derived
from $\mathbf{Y}$, and all arithmetic is over $\mathbb{F}_2$. The output
$\mathbf{Z} \in \{0,1\}^m$ is $\varepsilon$-close to uniform provided
both sources have sufficient independent min-entropy, with
$m \approx n \cdot k$ for min-entropy rate $k$.

The extractor output is partitioned into two fields:
\begin{equation}
  \begin{aligned}
    \qnum &\;=\; \mathrm{int}\!\Bigl(\mathbf{Z}\bigl[0:\lceil\log_2 1001\rceil\bigr],\, 2\Bigr)
            \bmod 1001, \\
    \mathbf{r} &\;=\; \mathbf{Z}\bigl[\lceil\log_2 1001\rceil :
                      \lceil\log_2 1001\rceil + 256\bigr],
  \end{aligned}
  \label{eq:partition}
\end{equation}
yielding $\qnum \in [0, 1000]$ and a 32-byte nonce~$\mathbf{r} \in \{0,1\}^{256}$
for replay resistance in the ToyLWE signature (Eq.~\ref{eq:seed}).

\begin{figure*}[t]
  \centering

\begin{tikzpicture}[
  font=\sffamily\small,
  >=Stealth
]


\node[font=\small\bfseries\sffamily, color=qpurple, anchor=west]
  at (0, 0.60) {$\bigstar$~(a) Circuit A: Two-Source QRNG};

\node[draw=qpurple, fill=qpurple!18, thick, rounded corners=4pt,
      minimum width=2.8cm, minimum height=0.60cm, align=center]
  (sv1) at (1.65, -0.15) {\small SV1 (ideal)};

\node[draw=qpurple, fill=qpurple!10, thick, rounded corners=4pt,
      minimum width=2.8cm, minimum height=0.50cm, align=center]
  (hm1) at (1.65, -1.05) {\small $H \to$ Measure};

\node[draw=qpurple, fill=qpurple!5, thick, rounded corners=4pt,
      minimum width=2.8cm, minimum height=0.45cm, align=center]
  (xout) at (1.65, -1.85) {\small $\mathbf{X} \in \{0,1\}^n$};

\draw[->, thick, draw=qpurple!60] (sv1) -- (hm1);
\draw[->, thick, draw=qpurple!60] (hm1) -- (xout);

\node[draw=qblue, fill=qblue!18, thick, rounded corners=4pt,
      minimum width=2.8cm, minimum height=0.60cm, align=center]
  (dm1) at (4.85, -0.15) {\small DM1 (noisy)};

\node[draw=qblue, fill=qblue!10, thick, rounded corners=4pt,
      minimum width=2.8cm, minimum height=0.50cm, align=center]
  (hm2) at (4.85, -1.05) {\small $H \to$ Measure};

\node[draw=qblue, fill=qblue!5, thick, rounded corners=4pt,
      minimum width=2.8cm, minimum height=0.45cm, align=center]
  (yout) at (4.85, -1.85) {\small $\mathbf{Y} \in \{0,1\}^n$};

\draw[->, thick, draw=qblue!60] (dm1) -- (hm2);
\draw[->, thick, draw=qblue!60] (hm2) -- (yout);

\node[draw=gray!50, fill=gray!12, thick, rounded corners=5pt,
      minimum width=5.4cm, minimum height=0.85cm, align=center]
  (extractor) at (3.25, -2.95)
  {\small \textbf{Toeplitz Two-Source Extractor}\\[-1pt]
   \scriptsize $\mathbf{Z} = T \cdot \mathbf{X} \oplus \mathbf{g}$};

\coordinate (inleft)  at ([xshift=-1.0cm]extractor.north);
\coordinate (inright) at ([xshift=+1.0cm]extractor.north);

\draw[->, thick, draw=black!45] (xout.south) -- ++(0,-0.18) -| (inleft);
\draw[->, thick, draw=black!45] (yout.south) -- ++(0,-0.18) -| (inright);

\node[draw=qpurple!50, fill=white, thick, rounded corners=4pt,
      minimum width=2.3cm, minimum height=0.45cm, align=center]
  (qnumout) at (1.80, -4.10) {\small $\qnum \in [0,1000]$};

\node[draw=qblue!50, fill=white, thick, rounded corners=4pt,
      minimum width=2.3cm, minimum height=0.45cm, align=center]
  (rout) at (4.70, -4.10) {\small $\mathbf{r} \in \{0,1\}^{256}$};

\coordinate (outcenter) at ([yshift=-0.20cm]extractor.south);
\draw[->, thick, draw=black!45] (extractor.south) -- (outcenter);
\draw[->, thick, draw=black!45] (outcenter) -| (qnumout.north);
\draw[->, thick, draw=black!45] (outcenter) -| (rout.north);

\node[font=\scriptsize\sffamily, align=left, anchor=west]
  at (0, -4.80) {
    Raw bit strings $\mathbf{X}$ (SV1) and $\mathbf{Y}$ (DM1) feed a
    Toeplitz extractor;\enspace output $\mathbf{Z}$ is partitioned into
    $q_{\mathrm{num}}$ and nonce~$\mathbf{r}$.
  };


\def\xoff{8.0}   

\def\yba{0}
\def\ybb{-0.85}

\node[font=\small\bfseries\sffamily, color=qblue, anchor=west]
  at (\xoff, 0.60)
  {$\bigstar$~(b) Circuit B: 2-qubit Bell State
   $|\Phi^+\rangle = \tfrac{1}{\sqrt{2}}(|00\rangle+|11\rangle)$};

\node[anchor=east, font=\small\sffamily]
  at (\xoff+0.35, \yba) {$q_0$};
\node[anchor=east, font=\small\sffamily]
  at (\xoff+0.35, \ybb) {$q_1$};

\draw[draw=black!70, thick]
  (\xoff+0.40, \yba) -- (\xoff+4.0, \yba);
\draw[draw=black!70, thick]
  (\xoff+0.40, \ybb) -- (\xoff+4.0, \ybb);

\node[draw=qblue, fill=qblue!18, thick,
      minimum width=0.75cm, minimum height=0.55cm,
      rounded corners=3pt, align=center,
      font=\small\sffamily] at (\xoff+1.1, \yba) {H};

\node[circle, fill=qblue, draw=qblue, inner sep=0pt,
      minimum size=6pt] at (\xoff+2.1, \yba) {};
\node[circle, draw=qblue, thick, inner sep=0pt,
      minimum size=12pt, font=\small] at (\xoff+2.1, \ybb) {$\oplus$};
\draw[draw=qblue, thick]
  (\xoff+2.1, \yba) -- (\xoff+2.1, \ybb);

\node[draw=black!70, fill=black!5, thick,
      minimum width=0.55cm, minimum height=0.55cm,
      rounded corners=3pt, font=\small\sffamily]
  at (\xoff+3.1, \yba) {M};
\node[draw=black!70, fill=black!5, thick,
      minimum width=0.55cm, minimum height=0.55cm,
      rounded corners=3pt, font=\small\sffamily]
  at (\xoff+3.1, \ybb) {M};

\node[font=\scriptsize\sffamily, align=left, anchor=west]
  at (\xoff, -1.55) {
    200 shots $\to$
    $\boldsymbol{\beta} = [P(|00\rangle),\,P(|01\rangle),\,
    P(|10\rangle),\,P(|11\rangle)]$
  };



\node[
  font=\small\sffamily, align=center,
  draw=gray!35, rounded corners=4pt, fill=gray!3,
  line width=0.6pt, inner sep=5pt
] at (12.0, -3.50) {
  {\color{qpurple}\rule{7pt}{7pt}}~H: Hadamard\quad
  {\color{qblue}$\bullet$--$\oplus$}: CNOT\quad
  {\color{gray!60}\rule{7pt}{7pt}}~Toeplitz\quad
  {\color{qpurple!40}\rule{6pt}{6pt}}~$\qnum$\,\,{\color{qblue!40}\rule{6pt}{6pt}}~$\mathbf{r}$\quad
  \fbox{\tiny\,M\,}~Measure
};

\end{tikzpicture}
\caption{Quantum circuits used in QSignAI.
    \textbf{(a) Circuit A} (two-source QRNG):
    Source~1~(SV1, ideal) and Source~2~(DM1, noisy) each execute
    single-qubit $H \to \text{measure}$ circuits; raw bit strings
    $\mathbf{X}$ and $\mathbf{Y}$ feed a Toeplitz two-source
    extractor producing uniform output $\mathbf{Z}$, partitioned
    into $\qnum \in [0,1000]$ and 32-byte nonce~$\mathbf{r}$.
    \textbf{(b) Circuit B} (Bell state $|\Phi^+\rangle$):
    {\color{qblue}$\blacksquare$}~H + CNOT create maximal entanglement;
    200 shots $\to$ $\bellvec=[P(00),\ldots,P(11)]$, encoding
    quantum entanglement in each participant's card colour.}
  \label{fig:circuits}
\end{figure*}

\subsubsection*{Two-Source Randomness Extraction}
\label{sec:extractor}

A single source of raw bits, even from a quantum simulator, is
insufficient for cryptographic-grade randomness: SV1 is a deterministic
state-vector simulator whose outcome distribution is fully predictable
given the same circuit, and taking the modal outcome of a simple
$H \to \text{measure}$ circuit would yield a deterministic value rather
than a genuinely random one. To overcome this, QSignAI employs a
\emph{two-source extractor}---a classical post-processing primitive that
combines two \emph{independent} weak random sources into a single output
that is statistically close to uniform.

The two-source model, formalised by Chor and Goldreich~\cite{ChorGoldreich1988} and
substantially developed by Dodis \emph{et al.}~\cite{dodis2004}, requires only
that each source contains sufficient min-entropy, with no assumption
that either source is fully uniform or that the sources are
uncorrelated beyond their independence. Formally, let $(\mathbf{X},\mathbf{Y})$
be two independent random variables on $\{0,1\}^n$ with min-entropies
$H_\infty(\mathbf{X}) \ge k$ and $H_\infty(\mathbf{Y}) \ge k$. A
$(n, k, \varepsilon)$ two-source extractor is a function
$\mathrm{Ext} : \{0,1\}^n \times \{0,1\}^n \to \{0,1\}^m$ such that
$\mathrm{Ext}(\mathbf{X}, \mathbf{Y})$ is $\varepsilon$-close to the
uniform distribution on $\{0,1\}^m$ in total variation distance.

\textbf{Toeplitz construction.} The specific construction used in
QSignAI is the Toeplitz-matrix extractor. A Toeplitz matrix $T$ has
constant diagonals ($T_{i,j} = t_{j-i}$), so an $m \times n$ matrix is
determined by only $m + n - 1$ bits. In our implementation, these
seed bits are drawn from the DM1 source $\mathbf{Y}$, while the SV1
source $\mathbf{X}$ provides the data vector. The bilinear form
$T \cdot \mathbf{X} \oplus \mathbf{g}$ (Eq.~\ref{eq:toeplitz}) is
computationally efficient---requiring only $\mathcal{O}(n \log n)$
operations via fast polynomial multiplication---and achieves the
information-theoretic extraction rate of the two-source model.

\textbf{Why this enables the full range.} With the two-source extractor
producing $\varepsilon$-uniform output bits, we obtain enough entropy to
select $\qnum$ uniformly from the full range $[0, 1000]$ (requiring
$\lceil \log_2 1001 \rceil = 10$ bits) and simultaneously derive the
32-byte nonce $\mathbf{r}$ (requiring 256 bits). Without extraction,
the raw single-source output would be limited to one bit per shot and
could not reliably produce 266 uniform bits. The extractor thus
\emph{simultaneously} enables (i)~uniform quantum randomness across the
entire required range and (ii)~replay resistance through the
nonce~$\mathbf{r}$ fed into the ToyLWE signature (Eq.~\ref{eq:seed}).

\subsection{Circuit B: 2-Qubit Bell State Measurement}

Circuit~B creates the Bell state
$|\Phi^+\rangle = (|00\rangle + |11\rangle)/\sqrt{2}$~\cite{bell1964,aspect1982}.
Two hundred measurement shots yield the probability distribution
$\bellvec = [P(|00\rangle),\,P(|01\rangle),\,P(|10\rangle),\,P(|11\rangle)]$.
In an ideal simulator, $P(|00\rangle) \approx P(|11\rangle) \approx 0.5$;
small deviations from ideal statistics provide a visual encoding of the simulated Bell-state statistics embedded in every participant's badge colour. The circuit parameters
for both circuits are summarised in Table~\ref{tab:circuits}.

\begin{table}[t]
  \caption{Quantum Circuit Parameters.
    {\color{qpurple}$\blacksquare$}~Circuit~A (Two-Source QRNG);
    {\color{qblue}$\blacksquare$}~Circuit~B (Bell state).}
  \label{tab:circuits}
  \centering\small
  \setlength{\tabcolsep}{4pt}
  \begin{tabularx}{\columnwidth}{l>{\raggedright\arraybackslash\columncolor{qpurple!8}}p{3.6cm}>{\raggedright\arraybackslash\columncolor{qblue!8}}X}
    \toprule
    \rowcolor{hdrpurple}
    \textcolor{white}{Parameter} &
    \textcolor{white}{Circuit~A (Two-Source QRNG)} &
    \textcolor{white}{Circuit~B (Bell)} \\
    \midrule
    Qubits        & 1 (SV1) + 1 (DM1)                & 2 \\
    Gate sequence & $H \to$ Measure ($\times$2 sources) & H, CNOT \\
    Shots         & $n$ per source                    & 200 \\
    Output        & $\qnum \in [0,1000]$, nonce~$\mathbf{r}$ & $\bellvec = [P(00),\dots,P(11)]$ \\
    Purpose       & Toeplitz extractor $\to$ uniform bits & Entropy + visual colour \\
    Exec.\ time   & 2--8~s (SV1 + DM1)                & 2--8~s (SV1) \\
    \bottomrule
    \multicolumn{3}{p{\dimexpr\columnwidth-8pt\relax}}{%
      \footnotesize\itshape
      Circuit~A runs two independent single-qubit $H$-measure circuits on
      SV1 (state-vector) and DM1 (density-matrix) simulators; raw bit
      strings $\mathbf{X}$, $\mathbf{Y}$ are condensed by a Toeplitz
      two-source extractor. Both circuits use OpenQASM~3.0.
      Results are written to S3 with prefix \textsf{amazon-braket-}.}
  \end{tabularx}
\end{table}

\begin{table*}[t]
  \caption{Entropy Budget for the QSignAI Signature Pipeline}
  \label{tab:entropy}
  \centering\small
  \begin{tabular}{llp{5.5cm}}
    \toprule
    \textbf{Component} & \textbf{Entropy} & \textbf{Notes} \\
    \midrule
    Two-source extractor (Circuit~A) & $\leq 10$~bits for $\qnum$ & Toeplitz extractor over SV1+DM1; $\varepsilon$-close to uniform \\
    Bell state $\bellvec$ (Circuit~B) & $\sim 0$~bits (security) & Visual encoding only; not key material \\
    Extractor nonce $\mathbf{r}$ & 256~bits & Dominant entropy source; replay resistance \\
    SHAKE-256 mixing & --- & Spreads entropy; adds none \\
    \midrule
    \textbf{Quantum-extracted share} & $\leq 10/266 \approx\! <4\%$ & Of total signature entropy \\
    \bottomrule
  \end{tabular}
\end{table*}

The quantum-extracted bits serve primarily as identity visualisation ($\qnum$ determines card hue via Eq.~(6)) and as a perceptual encoding of quantum phenomena, while the cryptographic strength of the signature comes from the 256-bit extractor-derived nonce and SHAKE-256 mixing.
\subsection{ToyLWE Signature Derivation}

The
quantum number seeds a ToyLWE-inspired signature using
SHAKE-256 (Secure Hash Algorithm~3 extendable-output function)~\cite{sha3}:
\begin{align}
  \mathscr{S}   &= \mathsf{SHAKE\text{-}256}(\mathscr{U}
                   \,\|\, \qnum \,\|\, r),
                   \label{eq:seed} \\
  \mathscr{H}   &= \mathsf{SHA\text{-}256}(\mathscr{S}[0{:}32])[0{:}12],
                   \label{eq:pkhash} \\
  \mathscr{G}   &= \mathsf{SHA\text{-}256}(\mathscr{H}_{\mathrm{msg}}
                   \,:\, \mathscr{H}_{\mathrm{ent}} \,:\, \mathscr{H}),
                   \label{eq:sig}
\end{align}
where $\mathbf{r}$ is the 32-byte nonce derived from the Toeplitz extractor output (Eq.~\ref{eq:partition}), ensuring replay resistance, $\mathscr{H}_{\mathrm{msg}}$
is the SHA-256 hash of the message content, and $\mathscr{H}_{\mathrm{ent}}$ is the
SHA-256 hash of $\qnum$. The public key hash $\pkh$
(12 uppercase hex characters) and 24-character base64 signature form the
badge displayed on each card: $\badge{452}{7B284BB3D413}$. The
structural analogy to LWE~\cite{regev2009} is that $\qnum$
plays the role of the secret and SHAKE-256 mixing plays the role of noise
addition. This is a demonstrative, pedagogical instantiation; the path to production is replacement with CRYSTALS-Dilithium (FIPS~204, ML-DSA)~\cite{fips204}.

\subsection{Visual Identity Encoding}

The quantum number and Bell state probabilities are mapped to a Hue-Saturation-Lightness~(HSL)
colour value:
\begin{align}
  h &= (\qnum \times 137.5) \bmod 360, \label{eq:hue} \\
  s &= 70 + \beta_0 \times 30,          \label{eq:sat} \\
  l &= 45 + \beta_3 \times 20,          \label{eq:light}
\end{align}
where $\beta_0 = P(|00\rangle)$ and $\beta_3 = P(|11\rangle)$. The golden-angle
increment (137.5\textdegree) maximises perceptual separation between
consecutive hue values, a technique from phyllotaxis. The degree of
quantum entanglement is encoded in the vividness and brightness of the
card colour: the colour of a card \emph{is} the quantum data, making
quantum science perceptually legible without any technical knowledge.

\subsection{Graceful Degradation}

A 30-second timeout or Braket task failure triggers a local
SHAKE-256-seeded fallback. Affected rows are flagged
\val{algorithm{:}\ "ToyLWE\text{-}local\text{-}fallback"} and
\val{device{:}\ "local\text{-}fallback"}. The wall never blocks; UX continuity
is preserved. The admin dashboard exposes the \api{device} field per
row, making quantum execution provenance auditable.


\subsection{Ethics and Data Governance}

The QSignAI deployment processes real participant data: Telegram usernames, message text, and optional photos. The following governance measures are in place.

\textit{Consent.} The \api{@mention} acts as a lightweight opt-in: only messages explicitly tagging the bot are processed. This follows the Telegram Bot API opt-in model and is disclosed to participants at event registration.

\textit{Data minimisation.} Only mention-bearing messages are persisted; all other group traffic is ignored. Photos are stored in an encrypted S3 bucket with 30-day automatic expiration. Username and message text are pseudonymised for analytics.

\textit{Retention and deletion.} Soft-delete (\api{hidden=true}) removes data from the public wall immediately. Hard deletion of all participant data from DynamoDB and S3 is available on request within 30 days, consistent with the right to erasure.

\textit{Moderation.} Event organisers have password-protected access to soft-delete inappropriate content. Bot Privacy Mode is disabled only for the specific group under active event management, not globally.

\textit{IRB.} No human subjects data has been collected to date; all testing has been conducted with bot-generated synthetic data and organiser self-testing. IRB approval must be obtained before any live participant data collection.
\section{User Experience Walkthrough}

The three-surface interface is shown in Fig.~\ref{fig:ux}. The interface
follows a three-surface guided flow.full screenshots of each surface are provided in Appendix~\ref{app:ux}.

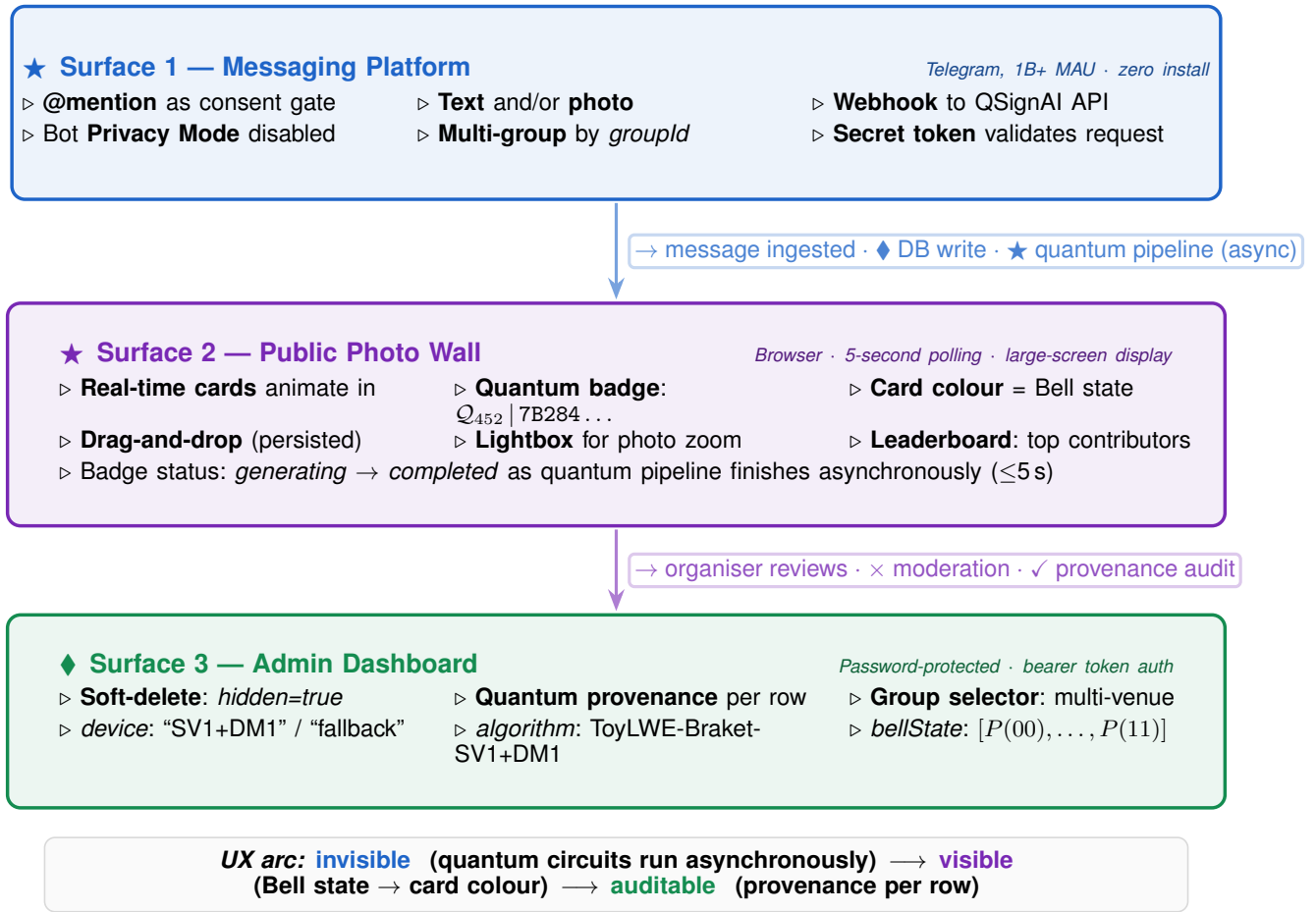
\begin{figure*}[!t]
  \centering

\begin{tikzpicture}[
  font=\sffamily,
  >=Stealth,
  every node/.style={inner sep=0pt, outer sep=0pt}
]

\node[font=\scriptsize\bfseries\sffamily, color=qblue,
      rotate=90, align=center]
  at (0.2, 0.0) {INPUT};
\node[font=\scriptsize\bfseries\sffamily, color=qpurple,
      rotate=90, align=center]
  at (0.2,-4.2) {DISPLAY};
\node[font=\scriptsize\bfseries\sffamily, color=qgreen,
      rotate=90, align=center]
  at (0.2,-8.2) {GOVERN};

\node[
  rounded corners=5pt, draw=qblue, fill=qblue!8, line width=1.3pt,
  minimum width=15.4cm, minimum height=2.6cm,
  align=left, text width=16.0cm, inner sep=4pt
] (s1) at (8.0, 0) {
  {\normalsize\bfseries\color{qblue}$\bigstar$\enspace
   Surface 1 --- Messaging Platform}%
  \hfill{\scriptsize\color{qblue!70!black}%
   \textit{Telegram, 1B+ MAU $\cdot$ zero install}}\par\vspace{4pt}
  {\small\begin{tabular}{@{}p{4.9cm}p{4.9cm}p{4.9cm}@{}}
    $\triangleright$~\textbf{@mention} as consent gate &
    $\triangleright$~\textbf{Text} and/or \textbf{photo} &
    $\triangleright$~\textbf{Webhook} to QSignAI API \\[2pt]
    $\triangleright$~Bot \textbf{Privacy Mode} disabled &
    $\triangleright$~\textbf{Multi-group} by \textsl{\textsf{groupId}} &
    $\triangleright$~\textbf{Secret token} validates request \\
  \end{tabular}}
};

\draw[->, line width=1.2pt, color=qblue!60]
  (8.0,-1.35) -- (8.0,-2.65)
  node[midway, right=5pt, font=\small\sffamily,
       color=qblue!80, fill=white, inner sep=2pt,
       rounded corners=2pt, draw=qblue!25]
  {$\rightarrow$~message ingested
   $\cdot$ $\blacklozenge$~DB write
   $\cdot$ $\bigstar$~quantum pipeline (async)};

\node[
  rounded corners=5pt, draw=qpurple, fill=qpurple!8, line width=1.3pt,
  minimum width=16.4cm, minimum height=3.0cm,
  align=left, text width=15.0cm, inner sep=4pt
] (s2) at (8.0,-4.2) {
  {\normalsize\bfseries\color{qpurple}$\bigstar$\enspace
   Surface 2 --- Public Photo Wall}%
  \hfill{\scriptsize\color{qpurple!70!black}%
   \textit{Browser $\cdot$ 5-second polling $\cdot$ large-screen display}}\par\vspace{4pt}
  {\small\begin{tabular}{@{}p{4.9cm}p{4.9cm}p{4.9cm}@{}}
    $\triangleright$~\textbf{Real-time cards} animate in &
    $\triangleright$~\textbf{Quantum badge}:
      $\mathcal{Q}_{452}\,|\,\mathtt{7B284\ldots}$ &
    $\triangleright$~\textbf{Card colour} = Bell state \\[2pt]
    $\triangleright$~\textbf{Drag-and-drop} (persisted) &
    $\triangleright$~\textbf{Lightbox} for photo zoom &
    $\triangleright$~\textbf{Leaderboard}: top contributors \\[2pt]
    \multicolumn{3}{@{}p{14.7cm}@{}}{%
      $\triangleright$~Badge status:
      \textsl{\textsf{generating}} $\to$ \textsl{\textsf{completed}}
      as quantum pipeline finishes asynchronously ($\leq$5\,s)} \\
  \end{tabular}}
};

\draw[->, line width=1.2pt, color=qpurple!60]
  (8.0,-5.75) -- (8.0,-6.85)
  node[midway, right=5pt, font=\small\sffamily,
       color=qpurple!80, fill=white, inner sep=2pt,
       rounded corners=2pt, draw=qpurple!25]
  {$\rightarrow$~organiser reviews
   $\cdot$ $\times$~moderation
   $\cdot$ $\checkmark$~provenance audit};

\node[
  rounded corners=5pt, draw=qgreen, fill=qgreen!8, line width=1.3pt,
  minimum width=16.4cm, minimum height=2.6cm,
  align=left, text width=15.0cm, inner sep=4pt
] (s3) at (8.0,-8.2) {
  {\normalsize\bfseries\color{qgreen}$\blacklozenge$\enspace
   Surface 3 --- Admin Dashboard}%
  \hfill{\scriptsize\color{qgreen!70!black}%
   \textit{Password-protected $\cdot$ bearer token auth}}\par\vspace{4pt}
  {\small\begin{tabular}{@{}p{4.9cm}p{4.9cm}p{4.9cm}@{}}
    $\triangleright$~\textbf{Soft-delete}:
      \textsl{\textsf{hidden=true}} &
    $\triangleright$~\textbf{Quantum provenance} per row &
    $\triangleright$~\textbf{Group selector}: multi-venue \\[2pt]
    $\triangleright$~\textsl{\textsf{device}}: ``SV1+DM1'' / ``fallback'' &
    $\triangleright$~\textsl{\textsf{algorithm}}: ToyLWE-Braket-SV1+DM1 &
    $\triangleright$~\textsl{\textsf{bellState}}:
      $[P(00),\ldots,P(11)]$ \\
  \end{tabular}}
};

\node[
  font=\small\bfseries\sffamily, align=center,
  draw=gray!30, rounded corners=4pt, fill=gray!4,
  line width=0.6pt, inner sep=5pt,
  minimum width=15.4cm, text width=15.0cm
] at (8.0,-10.4) {
  \textit{UX arc:}\enspace
  {\color{qblue}\textbf{invisible}}
  \enspace(quantum circuits run asynchronously)\enspace
  $\longrightarrow$\enspace
  {\color{qpurple}\textbf{visible}}
  \enspace(Bell state $\to$ card colour)\enspace
  $\longrightarrow$\enspace
  {\color{qgreen}\textbf{auditable}}
  \enspace(provenance per row)
};

\end{tikzpicture}
  \caption{Three-surface UX architecture of QSignAI.
    {\color{qblue}$\blacksquare$}~\textbf{\color{qblue}Surface~1}:
    bot-enabled messaging (Telegram, 1B+ MAU), \api{@mention}
    as consent gate;
    {\color{qpurple}$\blacksquare$}~\textbf{\color{qpurple}Surface~2}:
    public photo wall with quantum badge and Bell-state card colour;
    {\color{qgreen}$\blacksquare$}~\textbf{\color{qgreen}Surface~3}:
    admin dashboard with per-row quantum provenance audit.
    UX arc: {\color{qblue}invisible} $\to$
    {\color{qpurple}visible} $\to$ {\color{qgreen}auditable}.}
  \label{fig:ux}
\end{figure*}

\textit{Surface~1 (Messaging Platform):} participants use their existing
messaging client --- zero installation required. The \api{@mention}
acts simultaneously as a routing signal and a lightweight opt-in consent
gate. Only mention-bearing messages are persisted.

\textit{Surface~2 (Public Photo Wall):} the wall polls for new messages
every five seconds. Each card displays the sender's name, message text,
optional photo (click-to-zoom lightbox), and the quantum badge. Card
colour is derived from Bell state probabilities via
\eqref{eq:hue}--\eqref{eq:light}, making quantum entanglement visible to
a live audience. Participants drag cards to preferred positions; positions
are persisted as viewport percentages in DynamoDB, surviving window
resizes and refreshes. A leaderboard ranks top contributors by message
count.

\textit{Surface~3 (Admin Dashboard):} event organisers access a
password-protected dashboard exposing per-row quantum provenance
(\api{device}, \api{algorithm}, \api{signatureStatus},
\api{bellState}, \api{quantumNumber}). Soft-delete sets
\api{hidden\ =\ true} rather than removing rows, preserving the audit
trail and quantum signatures.

The UX arc moves from \emph{invisible} (quantum circuits running
asynchronously in the background) to \emph{visible} (Bell state
probabilities encoded as card colour on the public wall) to
\emph{auditable} (quantum execution provenance inspectable by
organisers). This arc is the AI-for-Science contribution: quantum
phenomena made legible to a live audience through a familiar social
interface.

The complete end-to-end interactive flow across all three surfaces is
shown in Fig.~\ref{fig:flow}, tracing a single participant's message
from submission through the quantum pipeline to the public wall and
admin dashboard.

\begin{figure*}[!t]
  \centering

\begin{tikzpicture}[
  font=\sffamily\small,
  >=Stealth,
  every node/.style={inner sep=0pt, outer sep=0pt}
]

\node[font=\small\bfseries\sffamily, align=center]
  at (2.0, 0) {$\bigstar$~Participant\\{\scriptsize(messaging platform)}};
\node[font=\small\bfseries\sffamily, align=center]
  at (9.0, 0) {$\star$~QSignAI System\\{\scriptsize(Bot $\cdot$ Cloud $\cdot$ Quantum)}};
\node[font=\small\bfseries\sffamily, align=center]
  at (15.5, 0) {$\blacklozenge$~Organiser\\{\scriptsize(Admin dashboard)}};

\draw[gray!25, dashed] (2.0, -0.5) -- (2.0,-20.0);
\draw[gray!25, dashed] (9.0, -0.5) -- (9.0,-20.0);
\draw[gray!25, dashed] (15.5,-0.5) -- (15.5,-20.0);

\node[
  rounded corners=4pt, draw=qblue, fill=qblue!10,
  line width=1.2pt, minimum width=4.5cm, minimum height=0.75cm,
  align=center, text width=4.2cm, inner sep=4pt
] (s1box) at (2.0,-1.3) {
  {\small\bfseries\color{qblue}$\bigstar$~Surface 1}\\
  {\scriptsize Messaging Platform $\cdot$ Telegram, 1B+ MAU}
};

\draw[->, thick, color=qblue]
  (2.0,-2.5) -- (9.0,-2.5)
  node[midway, above, font=\scriptsize\sffamily, color=qblue,
       fill=white, inner sep=1.5pt]
  {\textbf{1.}~\textit{@mention} message (text / photo)};

\node[
  rounded corners=4pt, draw=qgray, fill=qgray!8,
  line width=1.0pt, minimum width=9.5cm, minimum height=2.5cm,
  align=left, text width=9.1cm, inner sep=6pt
] (botbox) at (9.0,-4.1) {
  {\small\bfseries\color{qgray}$\star$~Bot Layer}\\[3pt]
  {\scriptsize
   \textbf{2.}~Webhook: validates secret token, detects \textit{@mention}, sanitizes text ($\leq$4096 chars)\\[2pt]
   \textbf{3.}~[If photo] Platform file API $\to$ download $\to$ S3 upload\\[2pt]
   \textbf{4.}~DB write Phase~1: \textit{signatureStatus} = ``generating'' $\to$ card on wall \textbf{immediately}}
};

\draw[->, thick, color=qgray, dashed]
  (9.0,-5.6) -- (2.0,-5.6)
  node[midway, below, font=\scriptsize\sffamily, color=qgray,
       fill=white, inner sep=1.5pt]
  {card visible (status: generating)};

\node[
  rounded corners=4pt, draw=qpurple, fill=qpurple!8,
  line width=1.0pt, minimum width=9.5cm, minimum height=5.4cm,
  align=left, text width=9.1cm, inner sep=6pt
] (qbox) at (9.0,-9.1) {
  {\small\bfseries\color{qpurple}$\bigstar$~Quantum Layer (async $\cdot$ non-blocking)}\\[4pt]
  {\scriptsize
   \textbf{5a.}~Circuit A — Two-source QRNG on SV1 + DM1 \quad(2--8\,s)\\[1pt]
   \quad SV1 (ideal): $H \to$ Measure $\to$ $\mathbf{X} \in \{0,1\}^n$\enspace
   $|$ DM1 (noisy): $H \to$ Measure $\to$ $\mathbf{Y} \in \{0,1\}^n$\\[1pt]
   \quad Toeplitz extractor: $\mathbf{Z} = T \cdot \mathbf{X} \oplus \mathbf{g}$\\[1pt]
   \quad $\mathbf{Z}$ partitioned into $q_{\mathrm{num}} \in [0,1000]$ and 32-byte nonce~$\mathbf{r}$\\[4pt]
   \textbf{5b.}~Circuit B — 2-qubit Bell State $|\Phi^+\rangle$ \quad(2--8\,s)\\[1pt]
   \quad $q_0{:}\ {-}H{-}\bullet{-}M$\enspace $q_1{:}\ {-}\oplus{-}M$\enspace
   200 shots $\to$ $[P(00),P(01),P(10),P(11)]$\\[4pt]
   \textbf{5c.}~ToyLWE derivation:\\[1pt]
   \quad $\mathsf{SHAKE\text{-}256}(\mathscr{U}\|q_{\mathrm{num}}\|\mathbf{r})$
   $\to$ $\mathcal{H}_{\mathrm{pk}}$ (12 hex chars)
   $\to$ $\mathcal{S}$ (24-char base64 signature)\\[1pt]
   \quad $\to$ $\mathrm{hsl}(h,s\%,l\%)$: hue from $q_{\mathrm{num}}$ (golden angle), sat/light from Bell state\\[4pt]
   \textbf{5d.}~DB write Phase~2: \textit{signatureStatus} = ``completed''}
};

\draw[->, thick, color=qpurple, dashed]
  (9.0,-12.0) -- (2.0,-12.0)
  node[midway, below, font=\scriptsize\sffamily, color=qpurple,
       fill=white, inner sep=1.5pt]
  {badge updates ($\leq$5\,s poll)};


\node[
  rounded corners=4pt, draw=qpurple, fill=qpurple!10,
  line width=1.2pt, minimum width=4.5cm, minimum height=0.75cm,
  align=center, text width=4.2cm, inner sep=4pt
] (s2box) at (2.0,-13.1) {
  {\small\bfseries\color{qpurple}$\bigstar$~Surface 2: Public Photo Wall}\\
  {\scriptsize Browser polls every 5\,s}
};

\node[
  rounded corners=4pt, draw=qgreen, fill=qgreen!10,
  line width=1.2pt, minimum width=4.5cm, minimum height=0.75cm,
  align=center, text width=4.2cm, inner sep=4pt
] (s3box) at (15.5,-13.1) {
  {\small\bfseries\color{qgreen}$\blacklozenge$~Surface 3: Admin Dashboard}\\
  {\scriptsize Password-protected $\cdot$ provenance audit}
};

\draw[->, thick, color=qgreen]
  (9.0,-12.1) -- (15.5,-12.1)
  node[midway, below, font=\scriptsize\sffamily, color=qgreen,
       fill=white, inner sep=1.5pt]
  {moderation request};

\node[
  rounded corners=3pt, draw=qpurple!50, fill=qpurple!5,
  line width=0.8pt, minimum width=4.3cm, minimum height=2.0cm,
  align=left, text width=4.0cm, inner sep=5pt
] at (2.0,-15.0) {
  {\scriptsize
   \textbf{Alice Chen}~\textit{@alice\_c}\\
   ``Hello from ACAIT 2026!''\\[2pt]
   $\bigstar$~$\mathcal{Q}_{452}\,|\,\mathtt{7B284BB3D413}$\\
   $\checkmark$~Quantum-authenticated}
};

\node[align=left, text width=4.3cm, font=\scriptsize\sffamily,
      inner sep=0pt]
  at (2.0,-17.0) {
  \textbf{6.}~Card renders with quantum badge\\[2pt]
  \textbf{7.}~Drag card $\to$ position persists for all viewers\\[2pt]
  \textbf{8.}~Click photo $\to$ fullscreen Lightbox\\[2pt]
  \textbf{9.}~Leaderboard ranks top contributors
};

\node[
  rounded corners=3pt, draw=qgreen!50, fill=qgreen!5,
  line width=0.8pt, minimum width=4.3cm, minimum height=4.0cm,
  align=left, text width=4.0cm, inner sep=5pt
] at (15.5,-16.0) {
  {\scriptsize
   \textbf{10.}~Login $\to$ Bearer token issued\\[2pt]
   \textbf{11.}~Review message table:\\
   \quad Alice: Q\#452, SV1+DM1 $\checkmark$\\
   \quad Bob: Q\#731, SV1+DM1 $\checkmark$\\
   \quad Carol: Q\#89, fallback $\triangle$\\[2pt]
   \textbf{12.}~Soft-delete: \textit{hidden=true}\\
   \quad (audit trail preserved)\\[2pt]
   \textbf{13.}~Provenance per row:\\
   \quad \textit{device}: ``SV1+DM1''\\
   \quad \textit{algorithm}: ToyLWE\\
   \quad \textit{bellState}: [.49,.01,.01,.49]}
};

\node[
  rounded corners=3pt, draw=gray!50, fill=gray!4,
  line width=0.8pt, minimum width=16.0cm,
  align=left, text width=15.6cm, inner sep=6pt
] at (8.75,-19.3) {
  {\small\bfseries Timing at a Glance}\quad
  {\scriptsize
  \begin{tabular}{@{}lll@{}}
    Steps 2--4   & $<1$\,s    & Webhook $\to$ DB write $\to$ card on wall \\
    Steps 5a--5d & 4--16\,s   & Quantum tasks (async, invisible to user) \\
    Step 6       & $\leq5$\,s & Next poll cycle $\to$ badge appears \\
    Steps 7--9   & instant    & Client-side interactions \\
    Fallback     & 30\,s      & Braket timeout $\to$ local crypto; wall unaffected \\
  \end{tabular}}
};

\end{tikzpicture}
  \caption{Complete interactive flow of QSignAI across three surfaces.
    \textbf{Actors:}
    $\bigstar$~Participant (messaging platform user);
    $\star$~QSignAI System (Bot, Cloud, Quantum);
    $\blacklozenge$~Organiser (Admin dashboard).
    \textbf{Surfaces:}
    {\color{qblue}$\blacksquare$}~Surface~1 (messaging platform, Telegram 1B+~MAU);
    {\color{qpurple}$\blacksquare$}~Surface~2 (public photo wall, 5-second polling);
    {\color{qgreen}$\blacksquare$}~Surface~3 (admin dashboard, provenance audit).
    Steps 2--4 complete in $<1$\,s; quantum tasks (Steps 5a--5d) run
    asynchronously in 4--16\,s; the badge updates on the next poll cycle
    ($\leq5$\,s). The fallback path (30\,s Braket timeout) preserves UX
    continuity. See Appendix~\ref{app:glossary} for term definitions.}
  \label{fig:flow}
\end{figure*}

\section{Deployment and Discussion}

\subsection{Answering RQ3: What Successful Deployment Demonstrates}

RQ3 asks whether a system combining quantum randomness and AI bot
mediation works in practice. The deployment answers this in three ways.
\textit{First, feasibility:} the system has been deployed on standard AWS
infrastructure, processed real event participants, and operated without
blocking on quantum task latency --- demonstrating that quantum circuits
are compatible with live event timing constraints when the pipeline is
designed asynchronously. \textit{Second, bidirectional value:} without
the quantum circuits, the bot's identity tokens are indistinguishable from
any PRNG-based system. Without the bot, the quantum circuits produce
outputs that no general audience ever sees. The combination creates a
system with a property neither component achieves alone. \textit{Third,
graceful degradation:} the fallback mechanism demonstrates that quantum
enhancement can be added to an AI system without introducing a single
point of failure.

\subsection{Deferred Measurable Comparisons}

The deployment establishes feasibility qualitatively. Measurable comparisons deferred to future work include: a rigorous NIST SP~800-90B~\cite{nist90b} comparison of quantum-randomness-seeded versus PRNG-seeded token distributions; systematic latency benchmarks across SV1, IonQ Aria, Rigetti Ankaa, and IBM Eagle; a controlled user study measuring whether exposure to quantum-derived visual badges builds quantum literacy; and a formal security comparison of ToyLWE versus CRYSTALS-Dilithium~\cite{fips204} under the LWE hardness assumption~\cite{regev2009}.

\subsection{Simulator Versus Physical QPU}

AWS Braket SV1 is a classical simulation of quantum behaviour --- its
randomness is ultimately pseudo-random at the hardware level. True quantum
randomness requires a physical quantum processing unit~(QPU). The architecture is QPU-ready: replacing the device Amazon Resource
Names~(ARNs) for SV1 and DM1 in the quantum signature module is the only
required change.The 2025 Nobel Prize
in Physics~\cite{martinis1985} recognised the hardware foundations that
make physical QPUs possible; integrating one is the primary near-term
extension.

\subsection{AI for Better Life Alignment}

QSignAI addresses the conference theme on three axes.
\textit{Accessibility:} quantum-authenticated identity is delivered through a channel over one billion people already use~\cite{durov2025} (addressable platform, not deployment scale), with no new hardware, no new application, and no technical knowledge required. \textit{Transparency:} every participant's identity has
verifiable quantum provenance, auditable by organisers through the admin
dashboard. \textit{Scientific literacy:} quantum concepts ---
superposition, entanglement, Bell states --- are made perceptually legible
to live audiences via visual badges, aligning with Sustainable Development
Goal~(SDG)~4 (Quality Education)~\cite{sdg2015}. The system also aligns
with SDG~9 (Industry, Innovation and Infrastructure), SDG~16 (Peace,
Justice and Strong Institutions), and SDG~17 (Partnerships for the Goals).

\section{Future Research Ecosystem}
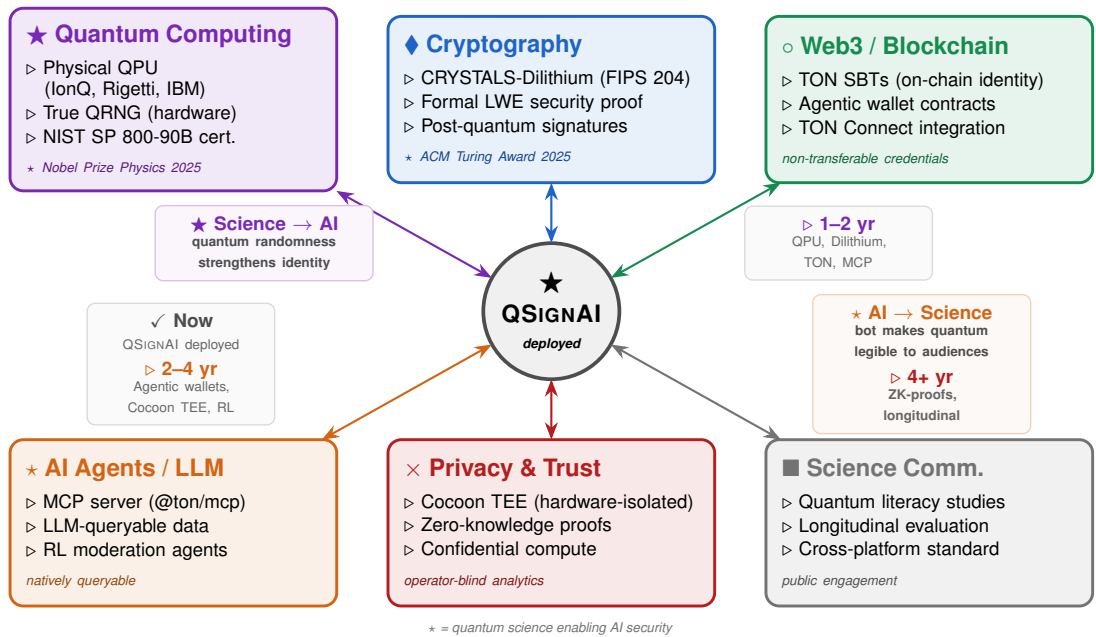
\begin{figure*}[!t]
  \centering

\begin{tikzpicture}[
  font=\sffamily\scriptsize,
  >=Stealth,
  every node/.style={inner sep=0pt, outer sep=0pt}
]


\node[
  rounded corners=5pt, draw=qpurple, fill=qpurple!10, line width=1.1pt,
  minimum width=4.2cm, minimum height=2.2cm,
  align=left, text width=3.9cm, inner sep=6pt
] (quantum) at (-5.0, 2.8) {
  {\small\bfseries\color{qpurple}$\bigstar$~Quantum Computing}\\[4pt]
  {\scriptsize
   $\triangleright$~Physical QPU\\
   \quad (IonQ, Rigetti, IBM)\\[1pt]
   $\triangleright$~True QRNG (hardware)\\[1pt]
   $\triangleright$~NIST SP~800-90B cert.\\[3pt]
   {\tiny\itshape\color{qpurple!70!black}$\star$~Nobel Prize Physics 2025}}
};

\node[
  rounded corners=5pt, draw=qblue, fill=qblue!10, line width=1.1pt,
  minimum width=4.2cm, minimum height=2.2cm,
  align=left, text width=3.9cm, inner sep=6pt
] (crypto) at (0.0, 2.8) {
  {\small\bfseries\color{qblue}$\blacklozenge$~Cryptography}\\[4pt]
  {\scriptsize
   $\triangleright$~CRYSTALS-Dilithium (FIPS 204)\\[1pt]
   $\triangleright$~Formal LWE security proof\\[1pt]
   $\triangleright$~Post-quantum signatures\\[3pt]
   {\tiny\itshape\color{qblue!70!black}$\star$~ACM Turing Award 2025}}
};

\node[
  rounded corners=5pt, draw=qgreen, fill=qgreen!10, line width=1.1pt,
  minimum width=4.2cm, minimum height=2.2cm,
  align=left, text width=3.9cm, inner sep=6pt
] (web3) at (5.0, 2.8) {
  {\small\bfseries\color{qgreen}$\circ$~Web3 / Blockchain}\\[4pt]
  {\scriptsize
   $\triangleright$~TON SBTs (on-chain identity)\\[1pt]
   $\triangleright$~Agentic wallet contracts\\[1pt]
   $\triangleright$~TON Connect integration\\[3pt]
   {\tiny\itshape\color{qgreen!70!black}non-transferable credentials}}
};

\node[
  circle, draw=black!70, fill=black!6, line width=1.4pt,
  minimum size=1.80cm, align=center,
  font=\small\bfseries\sffamily
] (core) at (0.0, 0.0) {
  {\normalsize$\bigstar$}\\[2pt]
  \textsc{QSignAI}\\
  {\tiny\itshape deployed}
};


\node[
  rounded corners=5pt, draw=qorange, fill=qorange!10, line width=1.1pt,
  minimum width=4.2cm, minimum height=2.2cm,
  align=left, text width=3.9cm, inner sep=6pt
] (agents) at (-5.0, -2.8) {
  {\small\bfseries\color{qorange}$\star$~AI Agents / LLM}\\[4pt]
  {\scriptsize
   $\triangleright$~MCP server (@ton/mcp)\\[1pt]
   $\triangleright$~LLM-queryable data\\[1pt]
   $\triangleright$~RL moderation agents\\[3pt]
   {\tiny\itshape\color{qorange!70!black}natively queryable}}
};

\node[
  rounded corners=5pt, draw=qred, fill=qred!10, line width=1.1pt,
  minimum width=4.2cm, minimum height=2.2cm,
  align=left, text width=3.9cm, inner sep=6pt
] (privacy) at (0.0, -2.8) {
  {\small\bfseries\color{qred}$\times$~Privacy \& Trust}\\[4pt]
  {\scriptsize
   $\triangleright$~Cocoon TEE (hardware-isolated)\\[1pt]
   $\triangleright$~Zero-knowledge proofs\\[1pt]
   $\triangleright$~Confidential compute\\[3pt]
   {\tiny\itshape\color{qred!70!black}operator-blind analytics}}
};

\node[
  rounded corners=5pt, draw=qscicomm, fill=qscicomm!10, line width=1.1pt,
  minimum width=4.2cm, minimum height=2.2cm,
  align=left, text width=3.9cm, inner sep=6pt
] (scicomm) at (5.0, -2.8) {
  {\small\bfseries\color{qscicomm}$\blacksquare$~Science Comm.}\\[4pt]
  {\scriptsize
   $\triangleright$~Quantum literacy studies\\[1pt]
   $\triangleright$~Longitudinal evaluation\\[1pt]
   $\triangleright$~Cross-platform standard\\[3pt]
   {\tiny\itshape\color{qscicomm!70!black}public engagement}}
};

\draw[<->, thick, color=qpurple] (core) -- (quantum);
\draw[<->, thick, color=qblue]   (core) -- (crypto);
\draw[<->, thick, color=qgreen]  (core) -- (web3);
\draw[<->, thick, color=qorange] (core) -- (agents);
\draw[<->, thick, color=qred]    (core) -- (privacy);
\draw[<->, thick, color=qscicomm](core) -- (scicomm);

\node[
  font=\scriptsize\bfseries\sffamily, align=center,
  draw=qpurple!30, rounded corners=3pt, fill=qpurple!4,
  line width=0.5pt, inner sep=4pt, text width=2.6cm
] at (-3.8, 0.9) {
  {\color{qpurple}$\bigstar$~Science $\to$ AI}\\
  {\tiny\color{black!70}quantum randomness\\strengthens identity}
};

\node[
  font=\scriptsize\sffamily, align=center,
  draw=gray!30, rounded corners=3pt, fill=gray!4,
  line width=0.5pt, inner sep=4pt, text width=2.2cm
] at (3.8, 0.9) {
  {\color{qpurple}$\triangleright$~\textbf{1--2 yr}}\\
  {\tiny\color{black!60}QPU, Dilithium,\\TON, MCP}
};

\node[
  font=\scriptsize\sffamily, align=center,
  draw=gray!30, rounded corners=3pt, fill=gray!4,
  line width=0.5pt, inner sep=4pt, text width=2.2cm
] at (-4.9, -0.7) {
  {\color{black!70}$\checkmark$~\textbf{Now}}\\
  {\tiny\color{black!60}\textsc{QSignAI} deployed}\\[2pt]
  {\color{qorange}$\triangleright$~\textbf{2--4 yr}}\\
  {\tiny\color{black!60}Agentic wallets,\\Cocoon TEE, RL}
};

\node[
  font=\scriptsize\bfseries\sffamily, align=center,
  draw=qorange!30, rounded corners=3pt, fill=qorange!4,
  line width=0.5pt, inner sep=4pt, text width=2.6cm
] at (4.9, -0.7) {
  {\color{qorange}$\star$~AI $\to$ Science}\\
  {\tiny\color{black!70}bot makes quantum\\legible to audiences}\\[2pt]
  {\color{qred}$\triangleright$~\textbf{4+ yr}}\\
  {\tiny\color{black!60}ZK-proofs,\\longitudinal}
};

\node[font=\tiny\itshape\sffamily, align=center, color=black!55]
  at (0.0, -4.2) {
  $\star$ = quantum science enabling AI security
};

\end{tikzpicture}
      \caption{Future research ecosystem with QSignAI as the deployed foundation, arranged as a 3$\times$2 grid of research domains. {\color{qpurple}$\blacksquare$}~\textbf{Quantum Computing} (physical QPU, hardware QRNG); {\color{qblue}$\blacksquare$}~\textbf{Cryptography} (CRYSTALS-Dilithium, formal LWE proof); {\color{qgreen}$\blacksquare$}~\textbf{Web3} (TON SBTs, agentic wallets); {\color{qorange}$\blacksquare$}~\textbf{AI Agents} (MCP server, RL moderation); {\color{qred}$\blacksquare$}~\textbf{Privacy} (Cocoon TEE, ZK-proofs); {\color{qscicomm}$\blacksquare$}~\textbf{Sci.\ Comm.} (quantum literacy studies). The timeline strip shows the four-horizon research roadmap.}

  \label{fig:ecosystem}
\end{figure*}

QSignAI is the deployed foundation for a broader research ecosystem,
illustrated in Fig.~\ref{fig:ecosystem}.

\textit{Near-term (1--2 years):} Physical QPU integration will replace SV1 and DM1 with independent
hardware quantum processors (IonQ Aria, Rigetti Ankaa, IBM Eagle) for genuine hardware quantum randomness, with NIST SP~800-90B certification~\cite{nist90b}. CRYSTALS-Dilithium replacement~\cite{fips204} will use the same entropy pipeline with production-grade PQC signatures and a formal security proof under the LWE hardness assumption~\cite{regev2009}. TON blockchain on-chain identity anchoring will issue Soulbound Tokens~(SBTs) as non-transferable Non-Fungible Token~(NFT) event credentials via TON Connect. A Model Context Protocol~(MCP) server~\cite{mcp2024} will expose quantum-authenticated participant data as an MCP resource, natively queryable by Large Language Model~(LLM) agents.

\textit{Mid-term (2--4 years):} Agentic wallet contracts will enable AI-initiated on-chain credential issuance. Cocoon Trusted Execution Environment~(TEE)-based confidential compute will support privacy-preserving analytics over participant data. Reinforcement Learning~(RL)~\cite{sutton2018} will power adaptive moderation agents.

\textit{Long-term (4+ years):} Zero-Knowledge~(ZK) proofs over post-quantum signatures, longitudinal quantum literacy studies, and cross-platform quantum identity standards represent the long-horizon research agenda.

\section{Conclusion}

QSignAI demonstrates that the relationship between AI and quantum science
is bidirectional: quantum randomness, rooted in the principles recognised
by the 2025 ACM Turing Award~\cite{bennett1984} and enabled by the
hardware recognised by the 2025 Nobel Prize in
Physics~\cite{martinis1985}, strengthens AI-driven identity (Science for
AI, RQ1); and the AI bot layer, following the template established by
AlphaFold2~\cite{jumper2021}, makes quantum phenomena accessible to
general audiences at event scale (AI for Science, RQ2). Neither direction
is theoretical --- both are demonstrated in a single production-deployed
system (RQ3). The central finding: a conversational AI bot and two quantum circuits on cloud simulators are sufficient to make quantum science tangible to over one billion potential users on the Telegram platform today~\cite{durov2025}.

\section*{Acknowledgment}

The authors thank the open-source community for contributions to the AWS
CDK, Next.js, and AWS Braket SDK ecosystems. The quantum circuit design
draws on the foundational work of Bennett and Brassard~\cite{bennett1984}
and the experimental verification of Bell inequalities by Aspect et
al.~\cite{aspect1982}. They also thank the participants and organizers
of the tutorials at The Web Conference 2026
\cite{10.1145/3774905.3793916} and IEEE ICBC 2026
\cite{guo2026blockchaininfrastructureintelligentcyberphysicalsocial}, where
\textit{QSignAI} was presented as an interactive demonstration and subsequently
refined based on participant feedback.

\bibliographystyle{IEEEtran}
\bibliography{references}

\appendix
\section{User Experience Screenshots}
\label{app:ux}


\section{User Experience Screenshots}
\label{app:ux}

The QSignAI user experience is designed around a three-surface
architecture that makes quantum science tangible to non-specialist
audiences through familiar social interfaces. The UX arc moves from
\emph{invisible} (quantum circuits running asynchronously in the
background) to \emph{visible} (Bell-state statistics encoded as card
colour on the public wall) to \emph{auditable} (quantum execution
provenance inspectable by organisers). Full screenshots of each
surface are shown below.

\begin{figure}[h]
  \centering
  \includegraphics[width=0.95\columnwidth]{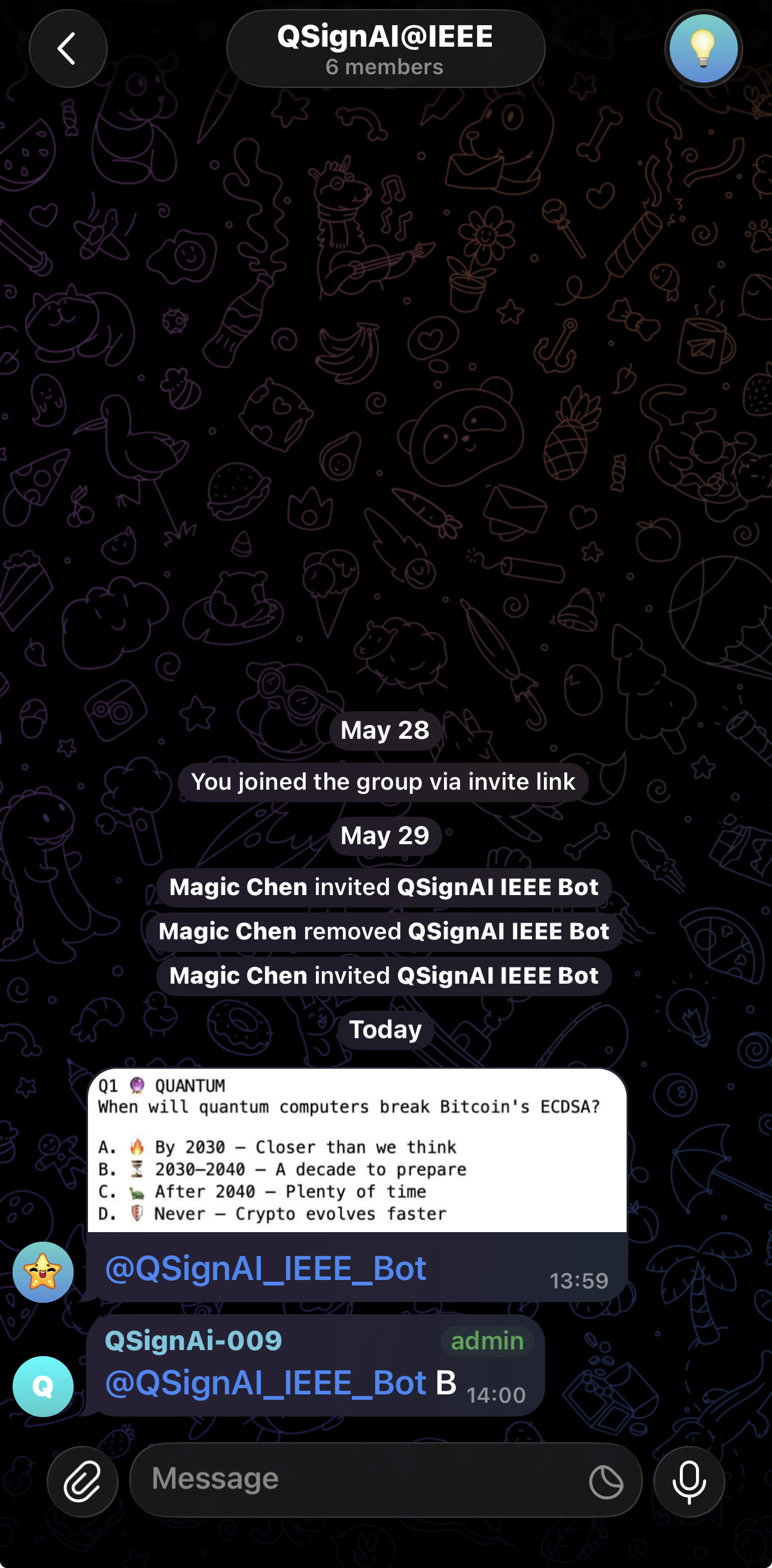}
  \caption{Surface 1 --- Messaging Platform (Telegram). The
    participant sends a message with \texttt{@mention} to the
    QSignAI bot, which acts as a lightweight opt-in consent gate.
    The bot acknowledges receipt and triggers the quantum pipeline
    asynchronously. No technical knowledge or additional software
    installation is required. The \texttt{@mention} simultaneously
    serves as a routing signal and a privacy opt-in, ensuring only
    intentionally tagged messages enter the system.}
  \label{fig:app:surface1}
\end{figure}

\paragraph*{Surface 1 details.}
As shown in Fig.~\ref{fig:app:surface1}, the participant interacts
through their existing Telegram client --- zero installation, zero
learning curve. The \texttt{@mention} tag routes the message to the
ECS Fargate backend via HTTPS webhook, where the bot validates the
secret token, sanitises text (4096-character maximum), detects the
mention entity, and writes an initial record to DynamoDB with
\textit{signatureStatus} = ``generating''. A card appears on the
public wall immediately (within $<1$\,s), even though the quantum
pipeline has not yet completed. If the message includes a photo,
the platform file API (20\,MB limit) downloads it and uploads to an
encrypted S3 bucket.

\begin{figure}[h]
  \centering
  \includegraphics[width=0.95\columnwidth]{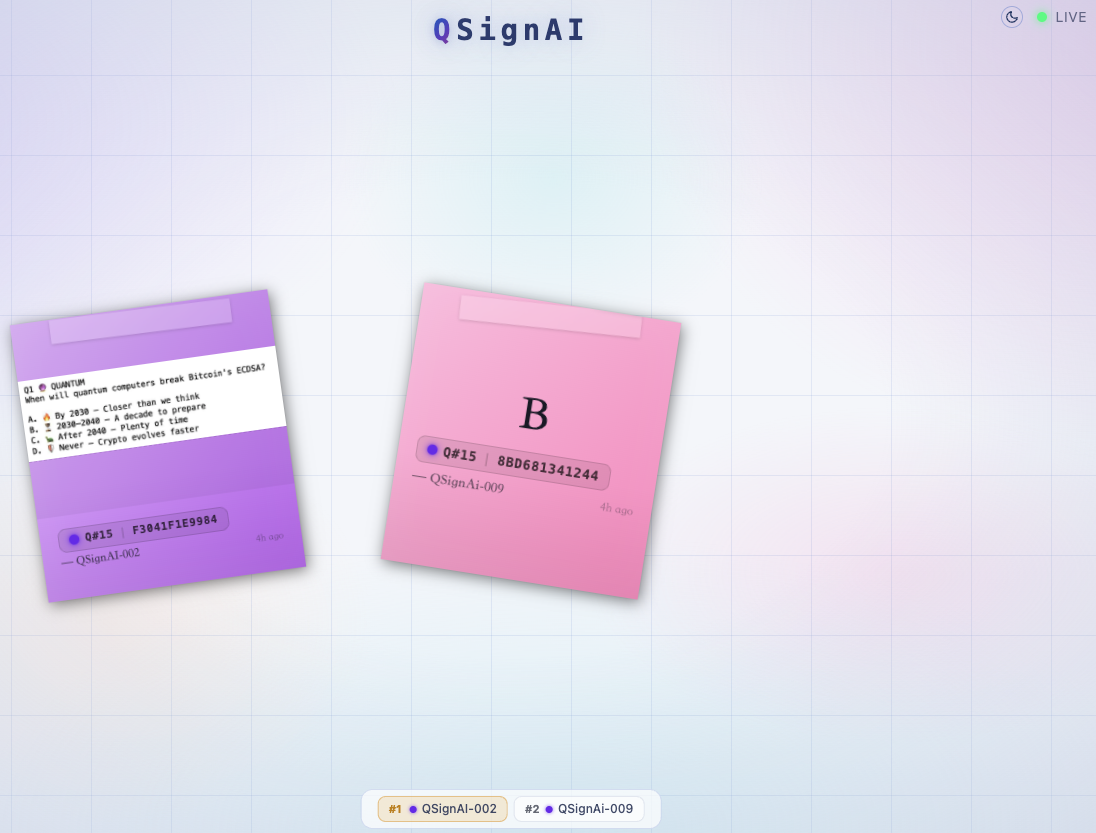}
  \caption{Surface 2 --- Public Photo Wall. Each card displays the
    sender's name, message text, optional photo (click-to-zoom
    lightbox), and the quantum badge
    $\mathcal{Q}_{\mathit{num}}\,|\,\mathtt{pkh}$. Card colour is
    derived from Bell-state probabilities via
    Eq.~\ref{eq:hue}--\ref{eq:light}, making quantum entanglement
    visible to a live audience without any technical knowledge.
    The wall polls for updates every five seconds.}
  \label{fig:app:surface2}
\end{figure}

\paragraph*{Surface 2 details.}
Fig.~\ref{fig:app:surface2} shows the public photo wall, the
central \emph{visible} layer of the UX arc. Each card renders with:
(1) sender name and Telegram handle; (2) message text; (3) optional
photo thumbnail that opens a fullscreen lightbox on click; (4) the
quantum badge in the form $\mathcal{Q}_{\mathit{num}}\,|\,
\mathtt{pkh}$ where $\mathit{num}$ is the quantum-randomness-seeded
identity number and \texttt{pkh} is the 12-character public key
hash; and (5) a card background colour derived from Bell-state
measurement statistics. The hue is computed via the golden-angle
increment (137.5\textdegree) from Eq.~\ref{eq:hue} to maximise
perceptual separation between consecutive cards. Saturation and
lightness come from $P(|00\rangle)$ and $P(|11\rangle)$
respectively (Eq.~\ref{eq:sat}--\ref{eq:light}), so the vividness
of the colour \emph{is} the quantum data. Participants can drag
cards to preferred positions; positions are persisted as viewport
percentages in DynamoDB, surviving window resizes and refreshes for
all viewers. A leaderboard ranks top contributors by message count.
The browser polls for new messages every five seconds, so the badge
updates from ``generating'' to ``completed'' asynchronously without
any user-visible blocking.

\begin{figure}[h]
  \centering
  \includegraphics[width=0.95\columnwidth]{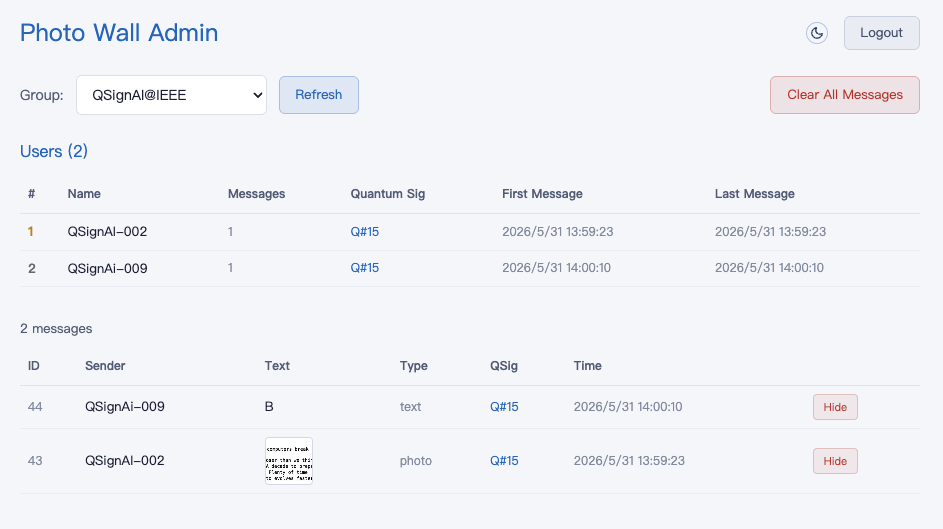}
  \caption{Surface 3 --- Admin Dashboard. Event organisers access
    a password-protected dashboard exposing per-row quantum
    provenance: \textit{device} (SV1+DM1 / fallback),
    \textit{algorithm}, \textit{signatureStatus},
    \textit{bellState} $= [P(|00\rangle),\ldots,P(|11\rangle)]$,
    and \textit{quantumNumber}. Soft-delete sets
    \texttt{hidden=true} preserving the full audit trail.}
  \label{fig:app:surface3}
\end{figure}

\paragraph*{Surface 3 details.}
Fig.~\ref{fig:app:surface3} presents the \emph{auditable} layer.
The password-protected admin dashboard exposes per-row quantum
provenance: \textit{device} (``SV1+DM1'' for two-source extractor
execution or ``fallback'' for local SHAKE-256),
\textit{algorithm} (``ToyLWE-Braket-SV1+DM1'' or
``ToyLWE-local-fallback''), \textit{signatureStatus}
(``generating'' / ``completed''),
\textit{bellState} $= [P(|00\rangle), P(|01\rangle),
P(|10\rangle), P(|11\rangle)]$, and \textit{quantumNumber}.
Soft-delete sets \texttt{hidden=true} rather than removing rows,
preserving the audit trail and quantum signatures indefinitely.
Organisers can also soft-delete inappropriate content. The
dashboard is bearer-token authenticated and includes a group
selector for multi-venue management.

\section{Glossary of Key Terms}
\label{app:glossary}

This appendix defines all key terms, abbreviations, and notation
used throughout the paper. Terms are grouped by domain; inline
definitions also appear at first use in the main text. Quantum
circuit notation follows OpenQASM~3.0; cryptographic notation
follows NIST FIPS standards.

\begin{table*}[h]
  \caption{Glossary of Key Terms, Abbreviations, and Notation
    (see also inline definitions at first use in the text).}
  \label{tab:glossary}
  \centering\small
  \setlength{\tabcolsep}{4pt}
  \begin{tabular}{lp{6.5cm}lp{6.5cm}}
    \toprule
    \rowcolor{hdrblue}
    \textbf{\textcolor{white}{Term}} &
    \textbf{\textcolor{white}{Definition}} &
    \textbf{\textcolor{white}{Term}} &
    \textbf{\textcolor{white}{Definition}} \\
    \midrule
    QRNG        & Quantum RNG: two-source extractor over independent quantum measurements &
    PRNG        & Pseudo-RNG: deterministic; theoretically reversible \\
    SV1         & AWS Braket state-vector simulator (ideal source) &
    DM1         & AWS Braket density-matrix simulator (noisy source) \\
    QPU         & Quantum Processing Unit: physical hardware &
    Toeplitz    & Toeplitz extractor: condenses two weak sources into uniform bits \\
    CNOT        & Controlled-NOT: 2-qubit entangling gate &
    Bell state  & $|\Phi^+\rangle=(|00\rangle+|11\rangle)/\sqrt{2}$ \\
    OpenQASM    & Open Quantum Assembly Language &
    ToyLWE      & Toy LWE: demonstrative lattice signature \\
    LWE         & Learning With Errors: lattice hard problem &
    PQC         & Post-Quantum Cryptography \\
    SHAKE-256   & SHA-3 extendable-output function &
    HSL         & Hue-Saturation-Lightness colour model \\
    $\qnum$     & Quantum number $\in[0,1000]$ from Circuit~A &
    $\bellvec$  & Bell state vector from Circuit~B \\
    $\mathscr{S}$ & Seed: $\mathsf{SHAKE\text{-}256}(\mathscr{U}\|\qnum\|\mathbf{r})$ &
    $\mathscr{H}$ & Public key hash: first 12 hex chars \\
    $\mathscr{G}$ & Signature: 24-char base64 string &
    $\mathscr{U}$ & Participant identity string (username) \\
    AWS         & Amazon Web Services: cloud platform &
    ECS         & Elastic Container Service \\
    DynamoDB    & Amazon managed NoSQL database &
    S3          & Amazon Simple Storage Service \\
    UX          & User Experience: three-surface design &
    MAU         & Monthly Active Users \\
    MCP         & Model Context Protocol~\cite{mcp2024} &
    LLM         & Large Language Model \\
    SBT         & Soulbound Token: non-transferable NFT &
    TEE         & Trusted Execution Environment \\
    ZK          & Zero-Knowledge proof &
    RL          & Reinforcement Learning~\cite{sutton2018} \\
    SDG         & Sustainable Development Goal~\cite{sdg2015} &
    nonce       & 32-byte replay-resistance value from extractor output \\
    \bottomrule
    \multicolumn{4}{p{14.0cm}}{%
      \footnotesize\itshape
      Terms grouped by domain. All circuit notation follows OpenQASM~3.0.
      Inline definitions at first use in the main text.}
  \end{tabular}
\end{table*}

\end{document}